\documentclass[pra,footinbib,twocolumn]{revtex4-1}
\def\Title{Critical dynamics at the Anderson localization mobility edge}

\usepackage[pdftex]{graphicx}
\usepackage{amsmath,amsfonts,amssymb,dsfont,color,bbm,cases}

\usepackage[colorlinks]{hyperref}
\newcommand{\avg}[1]{\overline{#1}}

\newcommand{\be}{\begin{equation}}
\newcommand{\ee}{\end{equation}}

\newcommand{\eps}{E}

\newcommand{\Ac}{A_\text{c}}
\newcommand{\Ec}{E_\text{c}}
\newcommand{\Dc}{D_\text{c}}
\newcommand{\kc}{k_\text{c}}
\newcommand{\lc}{l_\text{c}}
\newcommand{\tauc}{\tau_\text{c}}
\newcommand{\bk}{\mathbf{k}}
\newcommand{\bp}{\mathbf{p}}
\newcommand{\bq}{\mathbf{q}}
\newcommand{\br}{\mathbf{r}}

\newcommand{\bz}{\mathbf{z}}

\newcommand{\rmd}{\mathrm{d}}

\newcommand{\ket}[1]{\left|#1\right\rangle}

\newcommand{\els}{l}
\newcommand{\xpct}[1]{\left\langle#1\right\rangle}

\begin{document}
\author{Cord A.\ M\"uller$^1$, Dominique Delande$^2$, and Boris Shapiro$^3$}
\affiliation{$^1$Fachbereich Physik, Universit\"at Konstanz, 78457 Konstanz, Germany}
\affiliation{$^2$Laboratoire Kastler Brossel, UPMC-Sorbonne Universit\'es, CNRS,
ENS-PSL Research University, Coll\`ege de France, 4 Place Jussieu, 75005 Paris, France}
\affiliation{$^3$Department of Physics, Technion--Israel Institute of
  Technology, Haifa 32000, Israel} 

\title{\Title}
\begin{abstract} 
We study the critical dynamics of matter waves at the 3D Anderson mobility edge in cold-atom disorder quench experiments. 
General scaling arguments are supported by precision numerics for the spectral function, diffusion coefficient, and localization length in isotropic blue-detuned speckle potentials. We discuss signatures of critical slowdown in the time-dependent central column density of a spreading wave packet, and evaluate the prospects of observing anomalous diffusion right at criticality.    
\end{abstract} 

\maketitle

\section{Introduction}
 
Anderson localization refers to the remarkable phenomenon that configurational disorder can keep a phase-coherent system out of global equilibrium, with the eventual consequence that diffusive transport is entirely suppressed and thus resulting in an insulator \cite{Anderson1958,Abrahams2010}. 
Quantum systems 
show in dimensions  $d>2$ an Anderson transition \cite{Abrahams1979,Evers2008}, a critical change of transport properties from insulating to conducting that is rooted in the existence of a mobility edge $\Ec$, a critical point on  
the single-particle energy axis that separates localized from extended states. Since disordered electrons invariably interact \cite{Lee1985,Belitz1994}, making the unambiguous observation of the single-particle scenario very difficult, other physical carriers with much weaker interaction have been used to track the 3D Anderson transition: acoustic waves \cite{Hu2008}, light waves \cite{Wiersma1997,Stoerzer2006} (but see \cite{Sperling2016}), and cold atoms \cite{Modugno2010,Shapiro2012,Houches2009}. Notably, the universality and critical properties of the 3D Anderson transition have been thoroughly investigated with the quantum kicked rotor, a driven chaotic system, where localization operates in momentum space \cite{Chabe2008,Lemarie2009b}. Also cold-atom real-space experiments  \cite{Kondov2011,Jendrzejewski2012,Semeghini2015} have attempted to measure the mobility edge in spatially correlated laser speckle potentials. The basic idea of these wave-packet expansion experiments is that mobile atoms with energies $\eps>\Ec$ above the mobility edge escape, whereas localized atoms with energies $\eps<\Ec$ remain behind. From the measured localized fraction and initial energy distribution, one can then deduce the mobility edge. 

Concretely, let us assume that matter waves are prepared at time $t=0$ with an uncorrelated phase-space density $W(\bk,\br)=w(\bk)n_0(\br)$ \cite{Piraud2011,Jendrzejewski2012} and resulting energy distribution $A(E) = \int \rmd\bk A(\bk,E)w(\bk)$; the conditional probability $A(\bk,E)$ of a plane wave $\bk$ to have energy $E$ in the disorder potential is the spectral function.  The ensemble-averaged atom density at position $\br$ and time $t>0$ is
\be\label{density_n}
n(\br,t) =  \int \rmd\eps\, A(\eps)  \int\rmd\br_0 P(\eps,\br-\br_0,t) 
 n_0(\br_0),  
\ee
where $P(\eps,\br-\br_0,t)$ is the  
density (particle-hole) quantum propagator at energy $\eps$ from  $\br_0$ to $\br$ in time $t$ \cite{Akkermans2007,Evers2008,Kuhn2007}. 
The fraction of localized atoms, among all $N=\int \rmd\br_0n(\br_0)$ atoms initially present, then formally evaluates to 
\be
f_\text{loc} =\frac{1}{N}\int \rmd\br \lim_{t\to\infty}n(\br,t) = \int _{-\infty}^{\Ec} \rmd\eps\, A(\eps), 
\label{flocEc}
\ee 
where the last equality uses the long-time projection 
$ \int \rmd\br \lim_{t\to\infty}P(\eps,\br,t) = \Theta(\Ec-\eps)$ onto energies below the mobility edge. 
If the energy distribution $A(\eps)$ is known and covers the mobility edge, one can infer the position of $\Ec$ from the measured value $f_\text{loc}$. 

In an actual experiment, it is crucial to know how long one has to
wait for Eq.~\eqref{flocEc} to be valid, i.e., until the density of mobile atoms has 
dropped to zero in a given observation volume. 
Likewise, one has to ascertain carefully whether density profiles observed at finite times represent localized states, or rather comprise atoms that still diffuse, if only very slowly \cite{Skipetrov2008,McGehee2013,Mueller2014,McGehee2014}. 
Strictly speaking, the disappearance of the mobile fraction takes an infinite time since the diffusion coefficient becomes critically small near $\Ec$. A density measurement at a finite observation time then runs the danger of counting a certain, potentially sizeable, fraction of mobile atoms as localized, resulting  via Eq.~\eqref{flocEc} in an estimate for the mobility edge that is systematically too high. This effect may be one of the reasons why the experimental estimates for localized fractions and mobility edges of Refs.~\cite{Kondov2011,Jendrzejewski2012,Semeghini2015} are consistently above recent, accurate numerical estimates \cite{Delande2014}. Even if there existed a general awareness that critical dynamics come with diverging time and length scales, appropriate quantitative conclusions seem not to have been drawn until now. 
Finally, similar considerations should apply to the experimental characterization of the many-body localization transition \cite{Basko2006,Nandkishore2015,Altman2015,Schreiber2015,Bordia2016}, where subdiffusive transport is also expected to occur \cite{Agarwal2015,Potter2015,Vosk2015,BarLev2016,Choi2016,Luitz2016}. 

With this article, we wish to emphasize the principal, as well as practical relevance of critical dynamics around the 3D Anderson mobility edge in disorder-quench, matter-wave expansion experiments. 
This ana\-lysis significantly extends previous theories \cite{Kuhn2007,Skipetrov2008,Piraud2012,Piraud2013,Piraud2014} that used various versions of the self-consistent theory of localization, with largely uncontrolled approximations regarding the spectral function, critical exponents,  and position of the mobility edge. Indeed,  in Sec.~\ref{critical.sec} we illustrate our general arguments with precise numerical results for the spectral function, average density of states, diffusion coefficient, and localization length in 3D blue-detuned speckle disorder, as appropriate for present-day experiments. 
We do not attempt a quantitative comparison to one of the existing experiments \cite{Kondov2011,Jendrzejewski2012,Semeghini2015} since each of them uses its own particular preparation, measurement protocol and disorder configuration. 
Rather, our analysis relies on generic assumptions that represent the smallest common denominator for the existing cases, and it may thus serve as a conceptual guide rail for a more accurate analysis of experiments yet to come. 
Section \ref{column.sec} discusses the relevant time and length scales for the critical dynamics of a spreading wave packet. We focus our dicussion on the central column density, arguably a more trustworthy observable than small wing densities, and finally evaluate the prospects of observing anomalous diffusion right at criticality. 
Section~\ref{summary.sec} concludes.

\section{Critical energy and time scales} 
\label{critical.sec} 

\subsection{Anderson localization in 3D speckle disorder}

\label{speckle.sec}

Let $V(\br)$ denote a realization of a random, blue-detuned optical speckle potential \cite{Kuhn2007,Shapiro2012}. 
We assume that the disorder is statistically isotropic and homogeneous, leaving aside aspects such as 
anisotropy and finite size that may be relevant to particular experimental configurations.  The mean
$\overline{V(\br)}$ or ``sea level'' can be put to $0$ without loss of
generality, counting energies now from this level. The 
disorder strength is then defined by the variance 
$\overline{V(\br)^2} = V_0^2$. Higher-order moments 
of the single-point potential values $V=V(\br)$ are fully characterized by
the statistical distribution function $P(V)$. The blue-speckle distribution is strictly bounded from below and has the negative-exponential distribution 
$P(V) = \Theta(V+V_0)V_0^{-1} \exp[-(V+V_0)/V_0]$. 

Spatial correlations are captured by the covariance $\overline{V(\br)V(\br')} = V_0^2
C(\br-\br')$. 
For concreteness, we consider the Gaussian correlation $C(\br) = \exp[-\br^2/2\zeta^2]$. 
Such a correlation is \emph{generic}, 
defined by the property $\int\rmd \br C(\br) = \widetilde{C}(\bk=0)< \infty$, such that a white-noise description is applicable in the limit $k\zeta \ll 1$.
Experimentally, Gaussian correlation is relevant 
in the plane perpendicular to 
a focused laser beam with Gaussian intensity waist, and 
approximately isotropic 3D speckle potentials are created superposing 
several such patterns \cite{,Jendrzejewski2012,Semeghini2015} 
\footnote{Strictly speaking, 3D speckle potentials created by a superposition of monochromatic beams from ideal rectangular or circular apertures, for which a white-noise limit does not exist, are not generic in this sense \cite{Kuhn2007,Shapiro2012}. This caveat only affects the estimate \eqref{width} for the critical scale $W$, see \cite{Note3}, but does not invalidate our general analysis.   
}. 
The correlation length
defines the quantum correlation energy scale $E_\zeta =\hbar^2/m\zeta^2$. 

In the (semi-)classical limit $V_0\gg E_\zeta$ localization occurs very close to the
classical percolation threshold where $k\zeta \gg1$, such that
quantum interference effects are very small \cite{Shapiro2012}. We believe that expansion experiments in this regime \cite{Kondov2011,McGehee2013} have mainly probed diffusive dynamics \cite{Mueller2014,McGehee2014}. 
Here, we consider the opposite regime  
\be
\eta =\frac{V_0}{E_\zeta}\ll 1. 
\ee
 As a consequence, the potential
cannot produce locally bound states, and  
localization of matter waves becomes a quantum-mechanical,
multiple-scattering effect, known as \emph{Anderson localization}
(AL). 

It is by now established that the mobility edge $\Ec$ for blue-detuned speckle is located slightly below the 
sea level $\avg{V}=0$. 
This has been predicted by approximate, but sufficiently accurate treatments like the self-consistent
theory of localization \cite{Yedjour2010,Piraud2013} 
\footnote{A simple on-shell
approximation for the spectral function puts $\Ec$ above
sea-level \cite{Kuhn2007}, but this value is `red-shifted' below sea-level by the real part of the single-particle self-energy \cite{Piraud2013}.} 
and confirmed by exact numerical calculations 
\cite{Delande2014}. It also follows by a qualitative reasoning from 
the celebrated Ioffe-Regel 
criterion, which states that the Boltzmann description of classical transport has to be abandoned, and quantum effects may be expected to take over,  
once the wavelength becomes of the order of the mean-free path, $\kc \lc
\sim 1$. 
For the generic class of disorder and high energies well above sea level, the perturbation-theory mean-free path 
$\els \sim \zeta/\eta^2$ is independent of energy. Extrapolated to $l \sim \lc$, the Ioffe-Regel criterion then yields a characteristic energy scale $\hbar^2\kc^2/2m \sim W$  such that 
\footnote{The estimate \eqref{width} holds for generic disorder. In non-generic 3D speckle with an infrared divergence of the 3D pair correlator $\tilde C(\bk)$ \cite{Kuhn2007}, one rather has an energy-dependent mean free path 
$\els_\eps \sim k_\eps \zeta^2/\eta^2$, such that $W\sim \eta^2 E_\zeta = \eta V_0$, even larger than \eqref{width}.}
\be\label{width} 
W\sim \eta^4 E_\zeta = \eta^3V_0. 
\ee 
The mobility edge $\Ec$ itself then can be expected to lie near the energy that solves the
disorder-shifted dispersion relation for this momentum $\kc\sim \lc^{-1}$. 
Since the disorder shifts the bulk dispersion downward by the real part of the self-energy, $\Delta E \sim - \eta^2 E_\zeta$, which is larger in magnitude than $W$ for $\eta\ll 1$, the mobility edge finally ends up below sea level. 
We recall that the same scale $W$  is identified by a Lifshitz-tail argument coming from low energies \cite{Shapiro2012}. 
It is thus plausible to assume that $W$ is the single characteristic energy scale around $\Ec$, where quantum interference effects play a crucial role,  and thus yields an order of magnitude for the width of the critical interval. 
Although $W$ is rather small on both scales of $E_\zeta$ and $V_0$ in the regime of interest
$\eta\ll 1$, it can have a considerable impact on matter wave expansion dynamics, as shown in the following.

\subsection{Critical interval}
\label{criticalinterval.sec}

\begin{figure}
\includegraphics[width=0.9\linewidth]{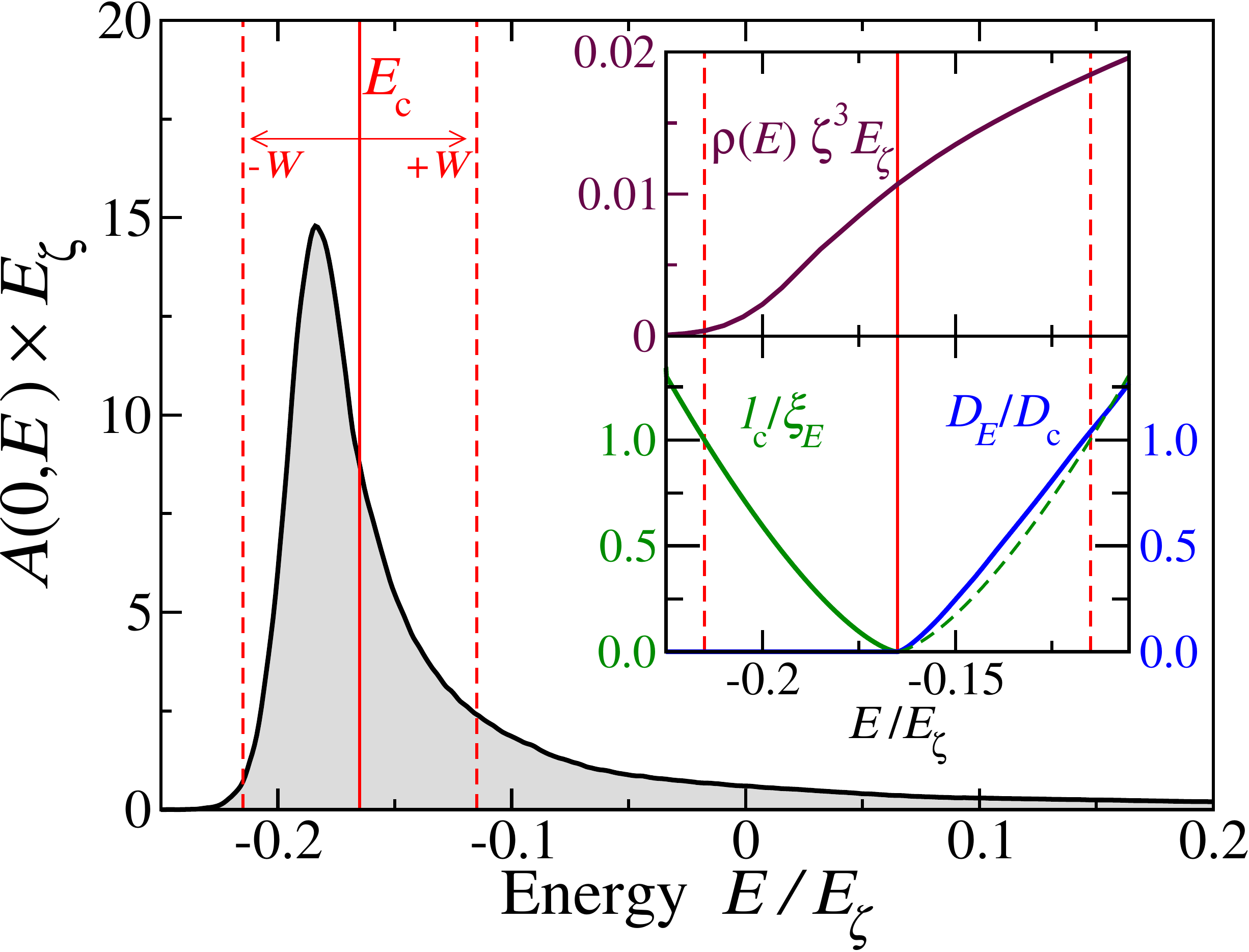}
\caption{
Main plot: Spectral function $A(0,E)$ of a matter wave with $\bk=0$ inside a blue-detuned laser speckle potential of zero mean and rms strength $V_0=0.5E_\zeta$; the unit of energy $E_\zeta=\hbar^2/m\zeta^2$ is set by a generic spatial Gaussian correlation length $\zeta$. The 3D Anderson mobility edge at $\Ec\approx-0.1652E_\zeta$ (red vertical line) lies below the `sea level' or potential mean at $E=0$.  The critical interval
with half width $W = 0.05E_\zeta$ (dashed vertical lines) contains initially more than two thirds of atoms; less than half of all atoms, with energy $E<\Ec$, 
will eventually be localized.    
Insets: Zoom in the critical interval with plots of the smooth average density of states $\rho(E)$, critical inverse localization length $\xi_E^{-1}$, and critical diffusion coefficient $D_E$, in units of $\lc$  and $\Dc=\hbar/3m$.  
}
\label{numplot.fig}
\end{figure} 

We proceed by presenting numerical data that prove the relevance of the critical interval around the 3D mobility edge. Our calculations use the single-particle Hamiltonian $H=\bp^2/2m + V(\br)$ of matter waves in blue-detuned speckle disorder with $\eta=V_0/E_\zeta=0.5$. 
Figure \ref{numplot.fig} shows the spectral
function $A(0,\eps)$ at zero momentum. 
The numerical routine propagates the initial state $\ket{\bk=0}$ in time with the Hamiltonian $H$ for each realization, followed by a Fourier transformation to energy and an ensemble average. 
The system size is chosen much larger than the scattering mean free
path, the spatial discretization much smaller than the correlation length $\zeta$, and the results are averaged over many disorder realizations, so that all results shown in Fig.~\ref{numplot.fig} have error bars
smaller than 1\% of their maximal value;
further details can be found in the appendix of Ref.~\cite{Trappe2015}.  

The spectral function thus obtained approximates the energy distribution $A(E)$ relevant for Eqs.~\eqref{density_n} and \eqref{flocEc} if the initial momentum distribution $w(\bk)$ is centered on $\bk=0$ and narrow enough. Indeed, 
the width of the spectral function $A(k,E)$ around $E=\Ec$ is of the order of $\lc^{-1}\sim \eta^2/\zeta$. If the width $\Delta k_0$ of the initial distribution $w(\bk_0)$ obeys $\Delta k_0\lc\ll1$, then $A(E) = \int \rmd\bk w(\bk) A(\bk,E) \approx A(0,E)$. Initial momentum distributions that are centered on finite $|\bk_0|>\lc^{-1}$ or that are much broader would provide a less optimal coverage of the mobility edge and will not be considered in the following. 

The upper inset of Fig.~\ref{numplot.fig} shows the average density of states per volume $\rho(E)$, which starts with an exponentially small Lifshitz tail \cite{Lifshitz1964,Lifshitz1988,Shapiro2012} from the exact lower bound of the spectrum  at $-V_0=-0.5E_\zeta$ (the combined lower bound of kinetic energy and centered blue speckle potential) and crosses over to the high-energy asymptotics $\propto E^{1/2}$ of the Galilean dispersion $E(\bp) = \bp^2/2m$ in 3D. An extrapolation of this bulk density of states to 0 gives an apparent lower band edge near $\Delta E \approx -0.20E_\zeta$.  

The spectral function and the average density of states are smooth functions of energy around the mobility edge
$\Ec$ that separates localized states
with $\eps<\Ec$ from extended states with $\eps>\Ec$ \cite{Delande2014,Pasek2015}. 
But transport coefficients   
show critical power-law behavior on both sides of $\Ec$, found here at $\Ec\approx-0.1652E_\zeta$, as shown in the lower inset of Fig.~\ref{numplot.fig}. The half width of the critical interval appears to be $W \approx 0.05E_\zeta$, in agreement with the estimate \eqref{width}. As the energy approaches the mobility edge $\Ec$ from below, the localization length diverges like 
\be\label{xiloc_crit}
\xi_\eps \sim \lc \left(\frac{W}{\Ec-\eps}\right)^{\nu},\qquad -W< \eps -\Ec <0,
\ee
with critical exponent 
$\nu\approx 1.58$  known only from numerical and laboratory experiments  \cite{Slevin1999,Lemarie2009b,Rodriguez2010,Lopez2012,Slevin2014}. 
The length $\lc$ is of the order of the elastic scattering length, a smooth function of energy at the transition. 
For our parameters, we find $\lc\approx4\zeta$, which complies with the Ioffe-Regel criterion $\kc\lc\sim 1$, with $\hbar\kc=\sqrt{2mW}$ the typical momentum at $\Ec$. 

As the energy approaches $\Ec$ from above, the diffusion coefficient vanishes as 
\be\label{D_crit} 
D_\eps \sim \Dc \left(\frac{\eps-\Ec}{W}\right)^s,\qquad 0 < \eps-\Ec<W.   
\ee
The critical exponents in Eqs.~\eqref{xiloc_crit} and \eqref{D_crit} are related by Wegner's law \cite{Wegner1976} $s=(d-2)\nu$, and thus $s=\nu$ in the present context. 
The diffusion coefficient reaches the ``quantum
unit of diffusion'' $\Dc=\hbar/3m$ around the energy $\Ec+W$ above the transition. 
We write $\Dc=\lc^2/\tau_c$ with $\tauc$ of the order of the elastic scattering time, also a smooth function of energy around $\Ec$. 

Numerically, we compute the critical quantities by a transfer-matrix calculation of the quasi-1D localization length $\xi_E(M)$ in a slab of transverse size $M$, extrapolated by finite-size scaling to the large-$M$ limit \cite{Kramer1993,Soukoulis1999}. 
Below $\Ec$, the 3D bulk localization length $\xi_E= \lim_{M\to\infty} \xi_E(M)$ is finite. Above $\Ec$, it is the localization length per cross section, ${\xi'_E}^{-1}= \lim_{M\to\infty} \xi_E(M)/M^2$, that is finite and plotted as the green dashed curve in the lower inset of Fig.~\ref{numplot.fig}. This length scale 
then gives the bulk diffusion coefficient as $D_E = [\pi\hbar\rho(E)\xi'_E]^{-1}$ \cite{Efetov1997}. One-parameter scaling actually constrains the critical exponents of $\xi_E$ and $\xi_E'$ to be equal, but since in our case $\rho(E)$ is not strictly constant over the full critical interval, $D_E$ does not obey the simple power law \eqref{D_crit} everywhere with the same $s=\nu$. Rather, Fig.~\ref{numplot.fig} displays a crossover for $D_E$ from a linear behavior with $s\approx1$ at higher energies toward the true $s=\nu$ close enough to $\Ec$, which is relevant for the long-time dynamics. 

For our showcase data of Fig.~\ref{numplot.fig}, the initial energy distribution $A(E)$ is composed of the  
regular localized fraction $f^\text{reg}_\text{loc} = A_{-\infty}^{\Ec-W} \approx 0.004$ 
[we note $A_a^b = \int_a^b \rmd E\,A(E)$],  
a critical localized fraction $f^\text{crit}_\text{loc} = A_{\Ec-W}^{\Ec} \approx 0.453$,
a critical diffusive fraction $f^\text{crit}_\text{diff} = A_{\Ec}^{\Ec+W}
    \approx 0.233$,
and a regular diffusive fraction $f^\text{reg}_\text{diff} =A_{\Ec+W}^\infty 
    \approx 0.310$. 
Clearly, for disorder strengths and energy distributions comparable to the above, we may expect the critical region to play an important role since it contains
a total critical fraction  of $f^\text{crit} = f^\text{crit}_\text{loc} + f^\text{crit}_\text{diff}
    \approx 0.686$, i.e.,  
more than two thirds of all atoms.

\subsection{Anomalous dynamics at finite times}

\begin{figure}[b]
\includegraphics[width=\linewidth]{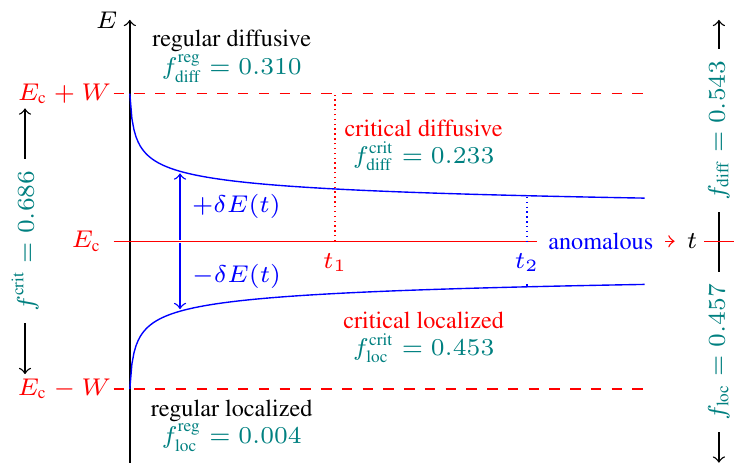}
\caption{Energy interval around the mobility edge $\Ec$ and different dynamical regimes. 
Numerical figures for the respective fractions are indicated for the showcase data of Fig.~\ref{numplot.fig}.   
The half width of the anomalous interval shrinks as $\delta\eps(t) = W (\tauc/t)^{1/3\nu}$, Eq.~\eqref{eps_of_t}. A spatial cutoff scale $\sigma_0$ (width of the initial wave packet if the central column density is monitored) introduces two crossover times: (1) The upper critical Thouless time $t_1 = \sigma_0^2/\Dc$, Eq.~\eqref{t1}, separates regular from critical diffusion. (2) The localization time $t_2=t_1\sigma_0/\lc$, Eq.~\eqref{t2}, separates critical from anomalous dynamics. 
}    
\label{dynamics_t.fig}
\end{figure} 

As the wave packet expands, the 
dynamical evolution progressively distinguishes between 
the total localized fraction $f_\text{loc} = f^\text{reg}_\text{loc} +  f^\text{crit}_\text{loc}
    \approx 0.457$
and the total diffusive fraction 
$f_\text{diff} = f^\text{reg}_\text{diff} +  f^\text{crit}_\text{diff}
    \approx 0.543$. 
Yet, 
in any experiment that probes dynamics of a wave
packet during a finite time $t$, it is impossible
      to sharply distinguish between localized and diffuse 
contributions near $\Ec$,  
since the resolution in energy is necessarily limited. Instead, energy components near criticality show 
a crossover behavior of anomalous diffusion \cite{Ohtsuki1997,Lemarie2010}, characterized by the sub\-diffusive law 
$\langle |x|\rangle \sim \lc (t/\tauc)^{1/3}$ that interpolates between the
truly diffusive and strongly localized regimes, where $\xpct{|x|} \sim l (t/\tau)^{1/2}$ and   $\xpct{|x|} \sim \xi t^0$, respectively.

One can estimate the half-width  $\delta\eps(t)$ of the anomalous energy interval by requiring that the critical 
localization length at energy $\Ec-\delta E$ be reached at time $t$ by anomalous
diffusion, $\xi_{\Ec-\delta E}=\lc(t/\tauc)^{1/3}$. Solving for $\delta E$ with Eq.~\eqref{xiloc_crit}, one
finds  
\be\label{eps_of_t}
\frac{\delta\eps(t)}{W} =  \left(\frac{\tauc}{t}\right)^{\frac{1}{3\nu}} =:\Delta(t). 
\ee 
Alternatively, this scale emerges by requiring that localization and diffusion cannot be distinguished, i.e., by equating $\xi_{\Ec-\delta E}$ with the critical diffusive radius at energy $\Ec+\delta E $ and time $t$. Solving $D_{\Ec+\delta E} t = \xi_{\Ec-\delta E}^2$ for $\delta E$ with the help of \eqref{xiloc_crit} and \eqref{D_crit} then yields \eqref{eps_of_t} as well.  

The anomalous energy interval shrinks as a function of time, 
as shown schematically in Fig.~\ref{dynamics_t.fig}, 
so that more and more particles are 
resolved as either localized or diffusive. But the anomalous energy interval shrinks very slowly, and only at infinitely long times does the 
separation between localized and mobile components become infinitely sharp. 
  
\subsection{Finite spatial resolution and crossover times} 

In real-life expansion experiments, the dynamics are limited not only temporally, but also spatially. For example, measurements of the total remaining numbers of particles \cite{Kondov2011,Semeghini2015} can only cover a finite observation range. Similarly, measurements of central densities \cite{Jendrzejewski2012} start from clouds with finite initial extension. Let us call $\sigma_0$ the corresponding spatial scale, and assume 
$\sigma_0\gg\lc$, which proves compatible with the assumption $\Delta k_0 \lc \ll 1 $ made for the momentum distribution in 
Sec.~\ref{criticalinterval.sec} above.   
The length $\sigma_0$ then introduces two characteristic crossover times that separate three different dynamical regimes, as schematically indicated in Fig.~\ref{dynamics_t.fig}. 

First, there is the `upper critical Thouless time' 
\be\label{t1} 
t_1 = \frac{\sigma_0^2}{\Dc} = \tauc \frac{\sigma_0^2}{\lc^2}, 
\ee 
namely the time required for the fastest critical, or slowest
regular,  diffusive atoms with diffusion coefficient $\Dc$ to explore the
scale $\sigma_0$. 
Starting with a broad enough energy distribution, the
early-time dynamics will be dominated by rapid, regular diffusion for $t\ll
t_1$, and cross over to slower, critical diffusion for $t\gg t_1$.  
As an order of magnitude, a value as large as $40\,\text{s}$ can be
realistic, taking the Palaiseau experiment \cite{Jendrzejewski2012} as
reference  where $\Dc\approx 0.25 \mu\text{m}^2/\text{ms}$ and
$\sigma_0\approx 100\,\mu\text{m}$ is the initial wave-packet size. 

Second, there is the time $t_2$ required for the slowest critical, or fastest anomalous, atoms with energy $\Ec+\delta E(t_2)$ to reach $\sigma_0$. Using \eqref{D_crit} and \eqref{eps_of_t}, this translates to the condition $\sigma_0^2=t_2\Dc \Delta(t_2)^\nu $. Equivalently, $t_2$ is the time where the localization length at $\Ec-\delta E(t_2)$ reaches $\sigma_0$. 
Both conditions agree on the `localization time' \cite{Skipetrov2008} 
\be\label{t2}
t_2  = \tauc \frac{\sigma_0^3}{\lc^3} = t_1 \frac{\sigma_0}{\lc},     
\ee 
or `watershed time' when the difference between localized and mobile components on the spatial scale $\sigma_0$ is resolved. 
From $t_2$ onwards, the localized components with $\xi_E<\sigma_0$ are essentially frozen, the diffusive components with $D_E>\sigma_0^2/t$ have essentially left, and the only remaining dynamics is due to the anomalous diffusion of a small fraction of particles with even larger critical localization lengths and smaller critically diffusive radii. 

In summary, one expects a double dynamical crossover \cite{Skipetrov2008}: first from fast, regular diffusion at times $t\ll t_1$ to slow, critical diffusion at times $t_1\ll t\ll t_2$ and then to anomalous behaviour for very long times $t\gg t_2$.

\section{Central column density}
\label{column.sec}

As a consequence of the previous considerations and associated long time scales, it becomes plausible that experiments tend to overestimate the localized fraction and thus yield values for $\Ec$ that are systematically too high. Clearly, one needs to extrapolate 
finite-time, finite-size data quite carefully in order to arrive at accurate estimates of the localized fraction. To facilitate this, we follow the general lines of \textcite{Skipetrov2008}, but instead of analyzing the sparse tails of expanding density distributions, we propose to monitor the decrease of the central column density at $x=y=0$, 
\be\label{defnperp}
n^\perp(t):=\int\rmd z\,n(0,0,z,t), 
\ee
from its initial value $n_0^\perp = n^\perp(0)$ as function of time. The central
column density, with its higher signal-to-noise ratio than the wing density,
has been used with success 
in dynamical localization experiments
\cite{Chabe2008,Lemarie2010}, and 
promises to be
useful in real space as well \cite{Jendrzejewski2012}. 

Whereas the localized component 
$ n^\perp_\text{loc} = f_\text{loc}n^\perp_0 $ 
remains essentially immobile from the start (to good approximation if $\lc\ll \sigma_0$ and thus $\xi_E \leq\sigma_0$ for the majority  of localized atoms), the mobile component will escape, and thus its contribution to the central column density will decrease in time. 

We consider a wave packet prepared in a system so large that ballistic modes are not populated. 
Let us also, in a first step, be deliberately oblivious of the anomalous dynamics and assume diffusion above $\Ec$ at all times by taking  $\delta E(t)=0$; the signatures of anomalous dynamics will be discussed in Sec.~\ref{anomalous.sec} below. 
In the diffusive interval $\eps>\Ec$, the density
propagator then is the Gaussian diffusion kernel 
\be
P(\eps,\br,t) = (4\pi D_\eps t)^{-\frac{3}{2}} \exp(-\br^2/4D_\eps t). 
\ee
Assuming an initial isotropic Gaussian distribution 
\be\label{n0Gauss}
n_0(\br)=\frac{N}{(2\pi\sigma_0^2)^{3/2}}\exp(-\br^2/2\sigma_0^2 ) ,  
\ee 
for which $n^\perp_0= N/2\pi\sigma_0^2$, 
 one finds from \eqref{density_n} and \eqref{defnperp} by Gaussian integration  
\be\label{nperpdiff} 
n^\perp(t) =  n^\perp_\text{loc} + N \int _{\Ec}^\infty \frac{\rmd\eps}{2\pi} \frac{A(\eps)}{\sigma_0^2+2D_\eps t}. 
\ee

If there were a finite time scale $t^\ast  = \max_\eps{\sigma_0^2/D_\eps}$ for all energies contained in the distribution $A(\eps)$ above $\Ec$, one would find for $t\gg t^\ast$ an asymptotic algebraic decrease like
\be\label{asympdiff}
n^\perp(t) \approx f_\text{loc}n^\perp_0 + \frac{N\langle
  D_\eps^{-1}\rangle}{4\pi t} +O(t^{-2})
\ee
where $\langle D_E^{-1}\rangle = \int _{\Ec}^\infty \rmd\eps A(\eps) D^{-1}_\eps$.   
A $1/t$ power-law fit could then reveal the constant offset and thus permit to extract $f_\text{loc}$, as in Ref.~\cite{Jendrzejewski2012}. However, this approach neglects the critical behaviour at the mobility edge. Indeed, whenever $A(\eps)$ is finite around $\Ec$, the  average of the inverse critical diffusion coefficient 
is ill-defined since 
\be
\langle D_\eps^{-1}\rangle \propto \int_{\Ec} \frac{\rmd\eps \,A(\eps)}{ (\eps-\Ec)^{\nu}} \sim \left.(\eps-\Ec)^{1-\nu}\right|_{\Ec} \to\infty. 
\ee 
It is actually impossible to enter the
supposed long-time regime $t\gg t^\ast$ because $\sigma_0^2/D_\eps\to\infty $ as
$\eps\to\Ec$, and the time required for the escape of all mobile atoms diverges as their energy approaches $\Ec$ from above. 

Of course, the argument leading to Eq.~\eqref{asympdiff} can be applied to the contribution of the \emph{regular diffusive} atoms with energies $E>\Ec+W$ and diffusion coefficients $D_E>\Dc$. For these atoms, $t^\ast = \sigma_0^2/\Dc=t_1$ is the upper critical Thouless time defined in Eq.~\eqref{t1}, 
and thus their contribution to \eqref{nperpdiff} vanishes like $t_1/t$ for $t\gg t_1$.

The regular diffusive atoms thus give way to the critically diffusive atoms with energies $\Ec<E<\Ec+W$ that take longer to disappear.
In the following, we discuss more quantitatively the contribution of critical dynamics to the central column density, dominant at times $t\gg t_1$, all the way to the anomalous diffusion at criticality that dominates at even longer times $t\gg t_2$.

\subsection{Critical diffusion}

 For the qualitative analysis of the expected critical dynamics, 
let us take the energy distribution $A(\eps)\approx
A(\Ec) =:\Ac$ to be constant
in the narrow critical interval above $\Ec$. There, the diffusion coefficient is $D_\eps = \Dc \Delta^\nu$, where $\Delta = (\eps-\Ec)/W\in [0,1]$ measures the distance to the critical
point. 
The critically diffusive atoms with energies $E\in[\Ec,\Ec+W]$ then contribute to the central column density \eqref{nperpdiff} with  
\be\label{densitycrit}
n^\perp_\text{diff} (t) \approx n^\perp_0 W\Ac \int _0
^1  
\frac{\rmd \Delta}{1+2(t/t_1) \Delta^\nu}. 
\ee
We only need to estimate the integral for times
$t \gg t_1=\sigma_0^2/\Dc$, when the regularly diffusive atoms have already 
disappeared. 
The integral is dominated by the behavior of the integrand near the lower bound 
and results in the algebraic decay 
\be\label{ndiffcrit}
n^\perp_\text{diff}(t) \sim n^\perp_0\left(\frac{t_1}{t}\right)^\frac{1}{\nu},\qquad t_1\ll t \ll t_2,
\ee
where constants of order unity, $W\!\Ac$ among them, are omitted. 
This power-law decay involving the critical exponent $\nu$ is much slower than the faster $t^{-1}$ decay of the regular diffusion, and consequently is quite easily mistaken for a saturation due to Anderson localization. 

\begin{figure}
\includegraphics[width=0.9\linewidth]{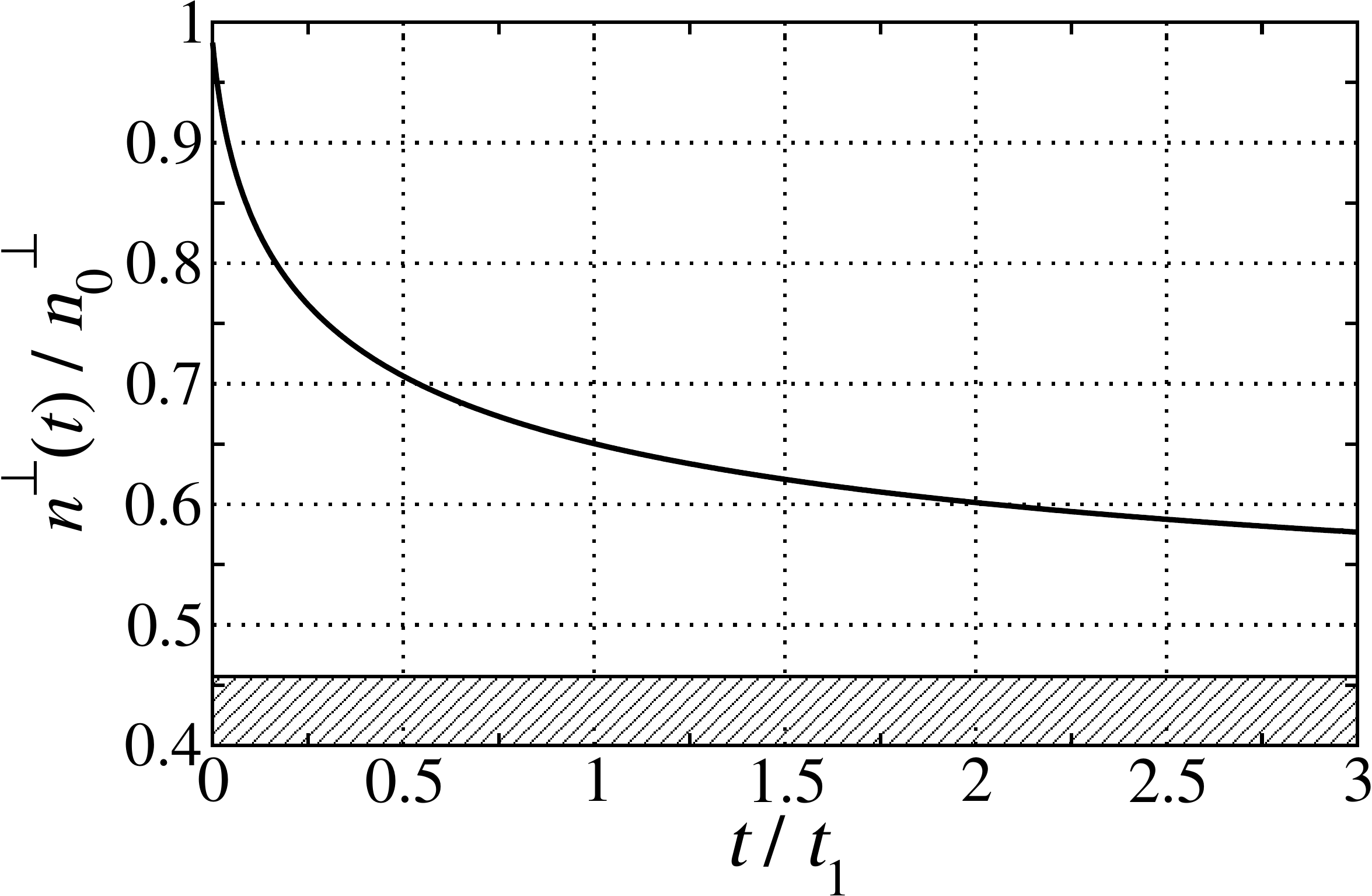}
\caption{%
Central column density, Eq.~\eqref{nperpdiff}, evaluated for the data of Fig.~\ref{numplot.fig}, as function of time. Around the upper critical Thouless time $t_1=\sigma_0^2/\Dc$, the fraction of regular diffusive atoms (here $f^\text{reg}_\text{diff}\approx 0.31$) has almost disappeared, and the slow critical diffusion of Eq.~\eqref{ndiffcrit} starts to dominate. Anomalous diffusion becomes only relevant at much longer times $t\gg t_2=\sigma_0 t_1/\lc$. Especially with noise on the data, one could be tempted to extrapolate the curve to a localized fraction of 0.52 or more; in reality, it is only $f_\text{loc}=0.457$, indicated by the shaded baseline. 
}
\label{nperpplot.fig}
\end{figure} 

To underscore this point, Fig.~\ref{nperpplot.fig} plots the central column density, Eq.~\eqref{nperpdiff}, as function of time for the parameters of Fig.~\ref{numplot.fig}. Around $t_1$, the fraction of regular diffusive atoms (here $f^\text{reg}_\text{diff}\approx 0.31$) has almost disappeared, and slow critical diffusion starts to dominate. 
It must be noted that the argument leading to the power-law prediction \eqref{ndiffcrit} slightly oversimplifies the actual situation. Since the spectral function $A(0,E)$ varies notably over the upper critical interval (see Fig.~\ref{numplot.fig}), the approximation \eqref{densitycrit} is not very accurate at early times of order $t_1$. Also, the caveat of \cite{Note1} applies such that the pure power law of Eq.~\eqref{ndiffcrit} is only reached rather slowly. Still, even if 
Fig.~\ref{nperpplot.fig} does not admit a global fit to the simple analytical estimate \eqref{ndiffcrit}, the overall critical slowdown is clearly visible. Especially if the data were noisy for longer times, a naive fit to a $t^{-1}$ power law would yield a measured localized fraction of 0.52 or more, significantly larger than its true value $f_\text{loc}=0.457$. 

How long does this critical decay last? Until here, we have neglected anomalous diffusion.  Reinserting the true lower bound of the critical diffusion interval at $\Ec+W\Delta(t)$, the integral to be evaluated in \eqref{densitycrit} is really only 
\be
 \int _{\Delta(t)}^1  
\frac{\rmd \Delta}{1+2(t/t_1) \Delta^\nu}. 
\ee
 The value of the integral for $t\gg t_1$ 
depends on the larger of the two cutoffs, either $1$ or $(t/t_1)\Delta(t)^\nu$. A crossover time $t_2$ between these two cases is defined by the condition
$\Delta(t_2)^\nu=t_1/t_2$, which 
yields Eq.~\eqref{t2}. 
At intermediate times $t_1\ll t \ll t_2$ such that $t_1/t$ is always greater than $\Delta(t)^\nu$, the lower integral bound can be set to zero; here the contribution of anomalous atoms is negligible, and the dynamics is indeed dominated by the critical diffusion of Eq.~\eqref{ndiffcrit}. 

At even longer times $t\gg t_2$, the integral in \eqref{densitycrit} is exhausted by
values close to its lower bound where $(t/t_1)\Delta(t)^\nu\gg 1$. With its upper bound sent to infinity, the integral then evaluates to $\Delta(t)^{1-\nu}/(\nu-1)$, and one obtains the long-time algebraic behaviour 
\be\label{ndiffcrit2}
n^\perp_\text{diff}(t) \sim n^\perp_0 \Delta(t_2) \left(\frac{t_2}{t}\right)^\frac{1+2\nu}{3\nu},\qquad t_2\ll t. 
\ee
Actually, this is the regime where anomalous diffusion becomes relevant, to be discussed next.

\subsection{Anomalous diffusion} 
\label{anomalous.sec}

Allowing for a frequency-dependent diffusion coefficient $D_\eps(\omega)$ 
in the Fourier-transformed diffusion kernel $P(E,\bq,\omega) \propto [D_E(\omega)\bq^2-i\omega]^{-1}$ \cite{Akkermans2007}, one can write the diffusion propagator 
\be\label{Pomega}
 P(\eps,\br,t)  = \int \frac{\rmd \omega}{2\pi} 
\frac{\exp\{-i\omega t-|\br|\sqrt{-i\omega/D_\eps(\omega)}\}}{4\pi D_\eps(\omega)|\br|}. 
\ee
Scaling theory and microscopic calculations 
\cite{Shapiro1982} have established that the frequency-dependent diffusion coeffcient $D_E(\omega)$ at the mobility edge is 
\be\label{Dcrit}
D_{\Ec}(\omega)= \Dc (-i\omega\tauc)^{1/3},  
\ee  
whence \eqref{Pomega} turns into the critical diffusion propagator 
\be\label{Pc}
P_\text{c}(\br,t) = \frac{1}{4\pi \Dc |\br|} \int \frac{\rmd \omega}{2\pi}
\frac{\exp\{-i\omega t-(-i\omega \tauc)^{1/3}|\br|/\lc\} }{(-i\omega\tauc)^{1/3}}.   
\ee
This expression, formally independent of energy, holds in the range $ \pm\delta E(t)=\pm W \Delta(t)$ around the mobility edge. Its contribution to the central column density  then is  
\be  
n^\perp_\text{c}(t) = 2 W\Ac \Delta(t) \int \rmd z\int\rmd \br_0 n_0(\br_0) P_\text{c}(|\hat\bz-\br_0|,t)
\ee
where $\hat\bz=(0,0,z)$. 
Using translation invariance 
$\int\rmd \br_0 n_0(\br_0) P(|\br-\br_0|)= \int\rmd \br_0 n_0(\br_0+\br) P(|\br_0|)$ and the initial distribution \eqref{n0Gauss}, we can perform the Gaussian integral over $z$ and thus face the task to evaluate 
\be  
n^\perp_\text{c}(t) = 2 n^\perp_0 W\Ac \Delta(t) \int\rmd \br_0 \exp\left\{-\frac{x_0^2+y_0^2}{2\sigma_0^2}\right\} P_\text{c}(|\br_0|,t)  
\ee
where the Gaussian restricts $x_0$ and $y_0$ to a few $\sigma_0$, but does not limit the range of $z_0$ anymore. 

We only need to consider long times $t\gg t _2 = \tauc \sigma_0^3/l^3$. Then the initial wave packet is much narrower than the kernel, and for all values $z_0$ not much larger than $\sigma_0$, we find 
that the integral \eqref{Pc} is dominated by the contribution from the pole
$\omega^{-1/3}$,  
\be  
\int \frac{\rmd \omega}{2\pi}
\frac{\exp\{-i\omega t\} }{(-i\omega\tauc)^{1/3}} = \frac{1}{\Gamma(1/3)(t^2\tauc)^{1/3}}
\ee 
with the well-known $t^{-2/3}$ behaviour \cite{Ohtsuki1997,Lemarie2009,Lemarie2010}. What about large excursions in $z_0\gg\sigma_0$? These are cut off by the anomalous propagator \eqref{Pc}, via a saddlepoint of the integral, 
$P_\text{c}(z_0\gg \sigma_0,t)\sim \exp\{ - a (\sigma_0^3t_2 /z_0^3t)^{1/2}\}$ (with $a$ of order
unity). The resulting large-$z_0$ cutoff at $\sigma_0(t/t_2)^{1/3}$ of the remaining  integral 
\be
\int_{\sigma_0}^{\sigma_0(t/t_2)^{1/3}}\frac{\rmd z_0}{z_0}  =  \frac{1}{3}\ln \frac{t}{t_2}
\ee
then provides a logarithmic correction, which leads to 
\be\label{nanomalous}
n^\perp_\text{c}(t) \sim n^\perp_0 
\Delta(t_2) \left(\frac{t_2}{t}\right)^\frac{1+2\nu}{3\nu}
\ln\frac{t}{t_2}. 
\ee
This result differs from the analogous estimate for the wings of density distributions \cite{Skipetrov2008}, where the anomalous contribution is predicted to decay for $t\gg t_2$ as the pure power law $t^{-(1+2\nu)/3\nu}$, i.e., algebraically more slowly than the critical contribution, which disappears as $t^{-(1+3\nu)/3\nu}$. In the present setting of the central column density, with its additional integral over $z$, the signature of anomalous dynamics is only the logarithmic correction \eqref{nanomalous} to the critical power-law decay \eqref{ndiffcrit2} with exponent $(1+2\nu)/3\nu\approx0.875$.

\section{Summary, Outlook} 
\label{summary.sec} 

In summary, we have discussed the specific challenges, and particular interest, 
of critical dynamics close to the 3D Anderson mobility edge in the expansion of noninteracting matter waves in strongly disordered laser speckle potentials. Given a certain spatial resolution $\sigma_0$, the critical Thouless time $t_1=\sigma_0^2/\Dc$ separates fast diffusion from a critical slowdown for dynamical observables. In particular, we find a crossover from a $t^{-1}$ decay of the central column density to a $t^{-1/\nu}$ decay.  
The signature of anomalous dynamics emerges 
after the `watershed time' 
$t_2 = t_1\sigma_0/\lc$, where all localized components on scales smaller than $\sigma_0$ are frozen and all faster diffusive components have left,  
as a logarithmic correction on top of an accelerated critical background decaying as $t^{-(1+2\nu)/3\nu}$. 

In our discussion, we have disregarded how exactly the localized
components with $\xi_E<\sigma_0$ contribute to the transient dynamics of the central column density at times $t\ll t_2$, since the diffusive components are expected to dominate.  
However, it is certainly interesting to study the dynamics of atoms that will eventually localize in the wake of a quantum quench. From momentum-space coherence signatures of Anderson localization in lower dimensions $d\leq 2$  \cite{Karpiuk2012,Lee2014,Ghosh2014,Micklitz2014,Micklitz2015}, we already know that the Heisenberg time of the localization
volume appears as a relevant time scale. 
An exhaustive description of critical localization dynamics close to the 3D mobility edge is an interesting challenge for future theoretical work. 

Also, we have neglected the possible impact of residual interactions. On a mean-field level, one expects that the localized fraction is replaced by a subdiffusive fraction, as discussed by Cherroret et al.\ in Ref.~\cite{Cherroret2014}, including the possibility of different critical exponents. However, quantitative conclusions for the real-space measurement scenarios discussed presently have not yet been drawn, to our knowledge, and thus remain to be investigated.

\begin{acknowledgments}
This work was granted access to the HPC resources of TGCC under the
allocation 2016-057644 made by GENCI (Grand Equipement National de
Calcul Intensif).
\end{acknowledgments}

\bibliography{\string~/Dropbox/Projects_Drop/projects,\string~/Dropbox/Bibs/ownpapers}

\begin{thebibliography}{66}%
\makeatletter
\providecommand \@ifxundefined [1]{%
 \@ifx{#1\undefined}
}%
\providecommand \@ifnum [1]{%
 \ifnum #1\expandafter \@firstoftwo
 \else \expandafter \@secondoftwo
 \fi
}%
\providecommand \@ifx [1]{%
 \ifx #1\expandafter \@firstoftwo
 \else \expandafter \@secondoftwo
 \fi
}%
\providecommand \natexlab [1]{#1}%
\providecommand \enquote  [1]{``#1''}%
\providecommand \bibnamefont  [1]{#1}%
\providecommand \bibfnamefont [1]{#1}%
\providecommand \citenamefont [1]{#1}%
\providecommand \href@noop [0]{\@secondoftwo}%
\providecommand \href [0]{\begingroup \@sanitize@url \@href}%
\providecommand \@href[1]{\@@startlink{#1}\@@href}%
\providecommand \@@href[1]{\endgroup#1\@@endlink}%
\providecommand \@sanitize@url [0]{\catcode `\\12\catcode `\$12\catcode
  `\&12\catcode `\#12\catcode `\^12\catcode `\_12\catcode `\%12\relax}%
\providecommand \@@startlink[1]{}%
\providecommand \@@endlink[0]{}%
\providecommand \url  [0]{\begingroup\@sanitize@url \@url }%
\providecommand \@url [1]{\endgroup\@href {#1}{\urlprefix }}%
\providecommand \urlprefix  [0]{URL }%
\providecommand \Eprint [0]{\href }%
\providecommand \doibase [0]{http://dx.doi.org/}%
\providecommand \selectlanguage [0]{\@gobble}%
\providecommand \bibinfo  [0]{\@secondoftwo}%
\providecommand \bibfield  [0]{\@secondoftwo}%
\providecommand \translation [1]{[#1]}%
\providecommand \BibitemOpen [0]{}%
\providecommand \bibitemStop [0]{}%
\providecommand \bibitemNoStop [0]{.\EOS\space}%
\providecommand \EOS [0]{\spacefactor3000\relax}%
\providecommand \BibitemShut  [1]{\csname bibitem#1\endcsname}%
\let\auto@bib@innerbib\@empty
\bibitem [{\citenamefont {Anderson}(1958)}]{Anderson1958}%
  \BibitemOpen
  \bibfield  {author} {\bibinfo {author} {\bibfnamefont {P.~W.}\ \bibnamefont
  {Anderson}},\ }\href {\doibase 10.1103/PhysRev.109.1492} {\bibfield
  {journal} {\bibinfo  {journal} {Phys. Rev.}\ }\textbf {\bibinfo {volume}
  {109}},\ \bibinfo {pages} {1492} (\bibinfo {year} {1958})}\BibitemShut
  {NoStop}%
\bibitem [{\citenamefont {Abrahams}(2010)}]{Abrahams2010}%
  \BibitemOpen
  \bibinfo {editor} {\bibfnamefont {E.}~\bibnamefont {Abrahams}},\ ed.,\ \href
  {http://www.worldscientific.com/worldscibooks/10.1142/7663} {\emph {\bibinfo
  {title} {50 Years of Anderson Localization}}}\ (\bibinfo  {publisher} {World
  Scientific},\ \bibinfo {year} {2010})\BibitemShut {NoStop}%
\bibitem [{\citenamefont {Abrahams}\ \emph {et~al.}(1979)\citenamefont
  {Abrahams}, \citenamefont {Anderson}, \citenamefont {Licciardello},\ and\
  \citenamefont {Ramakrishnan}}]{Abrahams1979}%
  \BibitemOpen
  \bibfield  {author} {\bibinfo {author} {\bibfnamefont {E.}~\bibnamefont
  {Abrahams}}, \bibinfo {author} {\bibfnamefont {P.~W.}\ \bibnamefont
  {Anderson}}, \bibinfo {author} {\bibfnamefont {D.~C.}\ \bibnamefont
  {Licciardello}}, \ and\ \bibinfo {author} {\bibfnamefont {T.~V.}\
  \bibnamefont {Ramakrishnan}},\ }\href {\doibase 10.1103/PhysRevLett.42.673}
  {\bibfield  {journal} {\bibinfo  {journal} {Phys. Rev. Lett.}\ }\textbf
  {\bibinfo {volume} {42}},\ \bibinfo {pages} {673} (\bibinfo {year}
  {1979})}\BibitemShut {NoStop}%
\bibitem [{\citenamefont {Evers}\ and\ \citenamefont
  {Mirlin}(2008)}]{Evers2008}%
  \BibitemOpen
  \bibfield  {author} {\bibinfo {author} {\bibfnamefont {F.}~\bibnamefont
  {Evers}}\ and\ \bibinfo {author} {\bibfnamefont {A.~D.}\ \bibnamefont
  {Mirlin}},\ }\href {\doibase 10.1103/RevModPhys.80.1355} {\bibfield
  {journal} {\bibinfo  {journal} {Rev. Mod. Phys.}\ }\textbf {\bibinfo {volume}
  {80}},\ \bibinfo {eid} {1355} (\bibinfo {year} {2008})}\BibitemShut {NoStop}%
\bibitem [{\citenamefont {Lee}\ and\ \citenamefont
  {Ramakrishnan}(1985)}]{Lee1985}%
  \BibitemOpen
  \bibfield  {author} {\bibinfo {author} {\bibfnamefont {P.~A.}\ \bibnamefont
  {Lee}}\ and\ \bibinfo {author} {\bibfnamefont {T.~V.}\ \bibnamefont
  {Ramakrishnan}},\ }\href {\doibase 10.1103/RevModPhys.57.287} {\bibfield
  {journal} {\bibinfo  {journal} {Rev. Mod. Phys.}\ }\textbf {\bibinfo {volume}
  {57}},\ \bibinfo {pages} {287} (\bibinfo {year} {1985})}\BibitemShut
  {NoStop}%
\bibitem [{\citenamefont {Belitz}\ and\ \citenamefont
  {Kirkpatrick}(1994)}]{Belitz1994}%
  \BibitemOpen
  \bibfield  {author} {\bibinfo {author} {\bibfnamefont {D.}~\bibnamefont
  {Belitz}}\ and\ \bibinfo {author} {\bibfnamefont {T.~R.}\ \bibnamefont
  {Kirkpatrick}},\ }\href {\doibase 10.1103/RevModPhys.66.261} {\bibfield
  {journal} {\bibinfo  {journal} {Rev. Mod. Phys.}\ }\textbf {\bibinfo {volume}
  {66}},\ \bibinfo {pages} {261} (\bibinfo {year} {1994})}\BibitemShut
  {NoStop}%
\bibitem [{\citenamefont {Hu}\ \emph {et~al.}(2008)\citenamefont {Hu},
  \citenamefont {Strybulevych}, \citenamefont {Page}, \citenamefont
  {Skipetrov},\ and\ \citenamefont {van Tiggelen}}]{Hu2008}%
  \BibitemOpen
  \bibfield  {author} {\bibinfo {author} {\bibfnamefont {H.}~\bibnamefont
  {Hu}}, \bibinfo {author} {\bibfnamefont {A.}~\bibnamefont {Strybulevych}},
  \bibinfo {author} {\bibfnamefont {J.~H.}\ \bibnamefont {Page}}, \bibinfo
  {author} {\bibfnamefont {S.~E.}\ \bibnamefont {Skipetrov}}, \ and\ \bibinfo
  {author} {\bibfnamefont {B.~A.}\ \bibnamefont {van Tiggelen}},\ }\href
  {\doibase 10.1038/nphys1101} {\bibfield  {journal} {\bibinfo  {journal} {Nat.
  Phys.}\ }\textbf {\bibinfo {volume} {4}},\ \bibinfo {pages} {945} (\bibinfo
  {year} {2008})}\BibitemShut {NoStop}%
\bibitem [{\citenamefont {Wiersma}\ \emph {et~al.}(1997)\citenamefont
  {Wiersma}, \citenamefont {Bartolini}, \citenamefont {Lagendijk},\ and\
  \citenamefont {Righini}}]{Wiersma1997}%
  \BibitemOpen
  \bibfield  {author} {\bibinfo {author} {\bibfnamefont {D.~S.}\ \bibnamefont
  {Wiersma}}, \bibinfo {author} {\bibfnamefont {P.}~\bibnamefont {Bartolini}},
  \bibinfo {author} {\bibfnamefont {A.}~\bibnamefont {Lagendijk}}, \ and\
  \bibinfo {author} {\bibfnamefont {R.}~\bibnamefont {Righini}},\ }\href
  {http://dx.doi.org/10.1038/37757} {\bibfield  {journal} {\bibinfo  {journal}
  {Nature}\ }\textbf {\bibinfo {volume} {390}},\ \bibinfo {pages} {671}
  (\bibinfo {year} {1997})}\BibitemShut {NoStop}%
\bibitem [{\citenamefont {St{\"o}rzer}\ \emph {et~al.}(2006)\citenamefont
  {St{\"o}rzer}, \citenamefont {Gross}, \citenamefont {Aegerter},\ and\
  \citenamefont {Maret}}]{Stoerzer2006}%
  \BibitemOpen
  \bibfield  {author} {\bibinfo {author} {\bibfnamefont {M.}~\bibnamefont
  {St{\"o}rzer}}, \bibinfo {author} {\bibfnamefont {P.}~\bibnamefont {Gross}},
  \bibinfo {author} {\bibfnamefont {C.~M.}\ \bibnamefont {Aegerter}}, \ and\
  \bibinfo {author} {\bibfnamefont {G.}~\bibnamefont {Maret}},\ }\href
  {\doibase 10.1103/PhysRevLett.96.063904} {\bibfield  {journal} {\bibinfo
  {journal} {Phys. Rev. Lett.}\ }\textbf {\bibinfo {volume} {96}},\ \bibinfo
  {eid} {063904} (\bibinfo {year} {2006})}\BibitemShut {NoStop}%
\bibitem [{\citenamefont {Sperling}\ \emph {et~al.}(2016)\citenamefont
  {Sperling}, \citenamefont {Schertel}, \citenamefont {Ackermann},
  \citenamefont {Aubry}, \citenamefont {Aegerter},\ and\ \citenamefont
  {Maret}}]{Sperling2016}%
  \BibitemOpen
  \bibfield  {author} {\bibinfo {author} {\bibfnamefont {T.}~\bibnamefont
  {Sperling}}, \bibinfo {author} {\bibfnamefont {L.}~\bibnamefont {Schertel}},
  \bibinfo {author} {\bibfnamefont {M.}~\bibnamefont {Ackermann}}, \bibinfo
  {author} {\bibfnamefont {G.~J.}\ \bibnamefont {Aubry}}, \bibinfo {author}
  {\bibfnamefont {C.~M.}\ \bibnamefont {Aegerter}}, \ and\ \bibinfo {author}
  {\bibfnamefont {G.}~\bibnamefont {Maret}},\ }\href
  {http://stacks.iop.org/1367-2630/18/i=1/a=013039} {\bibfield  {journal}
  {\bibinfo  {journal} {New J. Phys.}\ }\textbf {\bibinfo {volume} {18}},\
  \bibinfo {pages} {013039} (\bibinfo {year} {2016})}\BibitemShut {NoStop}%
\bibitem [{\citenamefont {Modugno}(2010)}]{Modugno2010}%
  \BibitemOpen
  \bibfield  {author} {\bibinfo {author} {\bibfnamefont {G.}~\bibnamefont
  {Modugno}},\ }\href {http://stacks.iop.org/0034-4885/73/i=10/a=102401}
  {\bibfield  {journal} {\bibinfo  {journal} {Rep. Progr. Phys.}\ }\textbf
  {\bibinfo {volume} {73}},\ \bibinfo {pages} {102401} (\bibinfo {year}
  {2010})}\BibitemShut {NoStop}%
\bibitem [{\citenamefont {Shapiro}(2012)}]{Shapiro2012}%
  \BibitemOpen
  \bibfield  {author} {\bibinfo {author} {\bibfnamefont {B.}~\bibnamefont
  {Shapiro}},\ }\href {http://stacks.iop.org/1751-8121/45/i=14/a=143001}
  {\bibfield  {journal} {\bibinfo  {journal} {J. Phys. A: Math. Theor.}\
  }\textbf {\bibinfo {volume} {45}},\ \bibinfo {pages} {143001} (\bibinfo
  {year} {2012})}\BibitemShut {NoStop}%
\bibitem [{\citenamefont {M\"uller}\ and\ \citenamefont
  {Delande}(2011)}]{Houches2009}%
  \BibitemOpen
  \bibfield  {author} {\bibinfo {author} {\bibfnamefont {C.~A.}\ \bibnamefont
  {M\"uller}}\ and\ \bibinfo {author} {\bibfnamefont {D.}~\bibnamefont
  {Delande}},\ }\enquote {\bibinfo {title} {Disorder and interference:
  localization phenomena},}\ in\ \href {\doibase
  10.1093/acprof:oso/9780199603657.003.0009} {\emph {\bibinfo {booktitle}
  {{Ultracold Gases and Quantum Information: Lecture Notes of the Les Houches
  Summer School in Singapore: Volume 91}}}},\ \bibinfo {editor} {edited by\
  \bibinfo {editor} {\bibfnamefont {C.}~\bibnamefont {Miniatura}}, \bibinfo
  {editor} {\bibfnamefont {L.-C.}\ \bibnamefont {Kwek}}, \bibinfo {editor}
  {\bibfnamefont {M.}~\bibnamefont {Ducloy}}, \bibinfo {editor} {\bibfnamefont
  {B.}~\bibnamefont {Gr\'{e}maud}}, \bibinfo {editor} {\bibfnamefont {B.-G.}\
  \bibnamefont {Englert}}, \bibinfo {editor} {\bibfnamefont {L.}~\bibnamefont
  {Cugliandolo}}, \bibinfo {editor} {\bibfnamefont {A.}~\bibnamefont {Ekert}},
  \ and\ \bibinfo {editor} {\bibfnamefont {K.~K.}\ \bibnamefont {Phua}}}\
  (\bibinfo  {publisher} {Oxford University Press},\ \bibinfo {year} {2011})\
  pp.\ \bibinfo {pages} {441--533},\ \Eprint {http://arxiv.org/abs/1005.0915}
  {arXiv:1005.0915} \BibitemShut {NoStop}%
\bibitem [{\citenamefont {Chab{\'e}}\ \emph {et~al.}(2008)\citenamefont
  {Chab{\'e}}, \citenamefont {Lemari{\'e}}, \citenamefont {Gr{\'e}maud},
  \citenamefont {Delande}, \citenamefont {Szriftgiser},\ and\ \citenamefont
  {Garreau}}]{Chabe2008}%
  \BibitemOpen
  \bibfield  {author} {\bibinfo {author} {\bibfnamefont {J.}~\bibnamefont
  {Chab{\'e}}}, \bibinfo {author} {\bibfnamefont {G.}~\bibnamefont
  {Lemari{\'e}}}, \bibinfo {author} {\bibfnamefont {B.}~\bibnamefont
  {Gr{\'e}maud}}, \bibinfo {author} {\bibfnamefont {D.}~\bibnamefont
  {Delande}}, \bibinfo {author} {\bibfnamefont {P.}~\bibnamefont
  {Szriftgiser}}, \ and\ \bibinfo {author} {\bibfnamefont {J.~C.}\ \bibnamefont
  {Garreau}},\ }\href {\doibase 10.1103/PhysRevLett.101.255702} {\bibfield
  {journal} {\bibinfo  {journal} {Phys. Rev. Lett.}\ }\textbf {\bibinfo
  {volume} {101}},\ \bibinfo {pages} {255702} (\bibinfo {year}
  {2008})}\BibitemShut {NoStop}%
\bibitem [{\citenamefont {Lemari{\'e}}\ \emph {et~al.}(2009)\citenamefont
  {Lemari{\'e}}, \citenamefont {Gr{\'e}maud},\ and\ \citenamefont
  {Delande}}]{Lemarie2009b}%
  \BibitemOpen
  \bibfield  {author} {\bibinfo {author} {\bibfnamefont {G.}~\bibnamefont
  {Lemari{\'e}}}, \bibinfo {author} {\bibfnamefont {B.}~\bibnamefont
  {Gr{\'e}maud}}, \ and\ \bibinfo {author} {\bibfnamefont {D.}~\bibnamefont
  {Delande}},\ }\href {http://stacks.iop.org/0295-5075/87/i=3/a=37007}
  {\bibfield  {journal} {\bibinfo  {journal} {Europhys. Lett.}\ }\textbf
  {\bibinfo {volume} {87}},\ \bibinfo {pages} {37007} (\bibinfo {year}
  {2009})}\BibitemShut {NoStop}%
\bibitem [{\citenamefont {Kondov}\ \emph {et~al.}(2011)\citenamefont {Kondov},
  \citenamefont {McGehee}, \citenamefont {Zirbel},\ and\ \citenamefont
  {DeMarco}}]{Kondov2011}%
  \BibitemOpen
  \bibfield  {author} {\bibinfo {author} {\bibfnamefont {S.~S.}\ \bibnamefont
  {Kondov}}, \bibinfo {author} {\bibfnamefont {W.~R.}\ \bibnamefont {McGehee}},
  \bibinfo {author} {\bibfnamefont {J.~J.}\ \bibnamefont {Zirbel}}, \ and\
  \bibinfo {author} {\bibfnamefont {B.}~\bibnamefont {DeMarco}},\ }\href
  {\doibase 10.1126/science.1209019} {\bibfield  {journal} {\bibinfo  {journal}
  {Science}\ }\textbf {\bibinfo {volume} {334}},\ \bibinfo {pages} {66}
  (\bibinfo {year} {2011})}\BibitemShut {NoStop}%
\bibitem [{\citenamefont {Jendrzejewski}\ \emph {et~al.}(2012)\citenamefont
  {Jendrzejewski}, \citenamefont {Bernard}, \citenamefont {M{\"u}ller},
  \citenamefont {Cheinet}, \citenamefont {Josse}, \citenamefont {Piraud},
  \citenamefont {Pezz{\'e}}, \citenamefont {Sanchez-Palencia}, \citenamefont
  {Aspect},\ and\ \citenamefont {Bouyer}}]{Jendrzejewski2012}%
  \BibitemOpen
  \bibfield  {author} {\bibinfo {author} {\bibfnamefont {F.}~\bibnamefont
  {Jendrzejewski}}, \bibinfo {author} {\bibfnamefont {A.}~\bibnamefont
  {Bernard}}, \bibinfo {author} {\bibfnamefont {K.}~\bibnamefont {M{\"u}ller}},
  \bibinfo {author} {\bibfnamefont {P.}~\bibnamefont {Cheinet}}, \bibinfo
  {author} {\bibfnamefont {V.}~\bibnamefont {Josse}}, \bibinfo {author}
  {\bibfnamefont {M.}~\bibnamefont {Piraud}}, \bibinfo {author} {\bibfnamefont
  {L.}~\bibnamefont {Pezz{\'e}}}, \bibinfo {author} {\bibfnamefont
  {L.}~\bibnamefont {Sanchez-Palencia}}, \bibinfo {author} {\bibfnamefont
  {A.}~\bibnamefont {Aspect}}, \ and\ \bibinfo {author} {\bibfnamefont
  {P.}~\bibnamefont {Bouyer}},\ }\href {http://dx.doi.org/10.1038/nphys2256}
  {\bibfield  {journal} {\bibinfo  {journal} {Nat. Phys.}\ }\textbf {\bibinfo
  {volume} {8}},\ \bibinfo {pages} {398} (\bibinfo {year} {2012})}\BibitemShut
  {NoStop}%
\bibitem [{\citenamefont {Semeghini}\ \emph {et~al.}(2015)\citenamefont
  {Semeghini}, \citenamefont {Landini}, \citenamefont {Castilho}, \citenamefont
  {Roy}, \citenamefont {Spagnolli}, \citenamefont {Trenkwalder}, \citenamefont
  {Fattori}, \citenamefont {Inguscio},\ and\ \citenamefont
  {Modugno}}]{Semeghini2015}%
  \BibitemOpen
  \bibfield  {author} {\bibinfo {author} {\bibfnamefont {G.}~\bibnamefont
  {Semeghini}}, \bibinfo {author} {\bibfnamefont {M.}~\bibnamefont {Landini}},
  \bibinfo {author} {\bibfnamefont {P.}~\bibnamefont {Castilho}}, \bibinfo
  {author} {\bibfnamefont {S.}~\bibnamefont {Roy}}, \bibinfo {author}
  {\bibfnamefont {G.}~\bibnamefont {Spagnolli}}, \bibinfo {author}
  {\bibfnamefont {A.}~\bibnamefont {Trenkwalder}}, \bibinfo {author}
  {\bibfnamefont {M.}~\bibnamefont {Fattori}}, \bibinfo {author} {\bibfnamefont
  {M.}~\bibnamefont {Inguscio}}, \ and\ \bibinfo {author} {\bibfnamefont
  {G.}~\bibnamefont {Modugno}},\ }\href {http://dx.doi.org/10.1038/nphys3339}
  {\bibfield  {journal} {\bibinfo  {journal} {Nat. Phys.}\ }\textbf {\bibinfo
  {volume} {11}},\ \bibinfo {pages} {554} (\bibinfo {year} {2015})}\BibitemShut
  {NoStop}%
\bibitem [{\citenamefont {Piraud}\ \emph {et~al.}(2011)\citenamefont {Piraud},
  \citenamefont {Lugan}, \citenamefont {Bouyer}, \citenamefont {Aspect},\ and\
  \citenamefont {Sanchez-Palencia}}]{Piraud2011}%
  \BibitemOpen
  \bibfield  {author} {\bibinfo {author} {\bibfnamefont {M.}~\bibnamefont
  {Piraud}}, \bibinfo {author} {\bibfnamefont {P.}~\bibnamefont {Lugan}},
  \bibinfo {author} {\bibfnamefont {P.}~\bibnamefont {Bouyer}}, \bibinfo
  {author} {\bibfnamefont {A.}~\bibnamefont {Aspect}}, \ and\ \bibinfo {author}
  {\bibfnamefont {L.}~\bibnamefont {Sanchez-Palencia}},\ }\href {\doibase
  10.1103/PhysRevA.83.031603} {\bibfield  {journal} {\bibinfo  {journal} {Phys.
  Rev. A}\ }\textbf {\bibinfo {volume} {83}},\ \bibinfo {pages} {031603}
  (\bibinfo {year} {2011})}\BibitemShut {NoStop}%
\bibitem [{\citenamefont {Akkermans}\ and\ \citenamefont
  {Montambaux}(2007)}]{Akkermans2007}%
  \BibitemOpen
  \bibfield  {author} {\bibinfo {author} {\bibfnamefont {E.}~\bibnamefont
  {Akkermans}}\ and\ \bibinfo {author} {\bibfnamefont {G.}~\bibnamefont
  {Montambaux}},\ }\href
  {http://www.loc.gov/catdir/enhancements/fy0803/2007279850-b.html} {\emph
  {\bibinfo {title} {{Mesoscopic physics of electrons and photons}}}}\
  (\bibinfo  {publisher} {Cambridge University Press},\ \bibinfo {address}
  {Cambridge},\ \bibinfo {year} {2007})\BibitemShut {NoStop}%
\bibitem [{\citenamefont {Kuhn}\ \emph {et~al.}(2007)\citenamefont {Kuhn},
  \citenamefont {Sigwarth}, \citenamefont {Miniatura}, \citenamefont
  {Delande},\ and\ \citenamefont {M{\"u}ller}}]{Kuhn2007}%
  \BibitemOpen
  \bibfield  {author} {\bibinfo {author} {\bibfnamefont {R.}~\bibnamefont
  {Kuhn}}, \bibinfo {author} {\bibfnamefont {O.}~\bibnamefont {Sigwarth}},
  \bibinfo {author} {\bibfnamefont {C.}~\bibnamefont {Miniatura}}, \bibinfo
  {author} {\bibfnamefont {D.}~\bibnamefont {Delande}}, \ and\ \bibinfo
  {author} {\bibfnamefont {C.~A.}\ \bibnamefont {M{\"u}ller}},\ }\href
  {http://www.iop.org/EJ/article/1367-2630/9/6/161/njp7_6_161.pdf} {\bibfield
  {journal} {\bibinfo  {journal} {New J. Phys.}\ }\textbf {\bibinfo {volume}
  {9}},\ \bibinfo {pages} {161} (\bibinfo {year} {2007})}\BibitemShut {NoStop}%
\bibitem [{\citenamefont {Skipetrov}\ \emph {et~al.}(2008)\citenamefont
  {Skipetrov}, \citenamefont {Minguzzi}, \citenamefont {van Tiggelen},\ and\
  \citenamefont {Shapiro}}]{Skipetrov2008}%
  \BibitemOpen
  \bibfield  {author} {\bibinfo {author} {\bibfnamefont {S.~E.}\ \bibnamefont
  {Skipetrov}}, \bibinfo {author} {\bibfnamefont {A.}~\bibnamefont {Minguzzi}},
  \bibinfo {author} {\bibfnamefont {B.~A.}\ \bibnamefont {van Tiggelen}}, \
  and\ \bibinfo {author} {\bibfnamefont {B.}~\bibnamefont {Shapiro}},\ }\href
  {\doibase 10.1103/PhysRevLett.100.165301} {\bibfield  {journal} {\bibinfo
  {journal} {Phys. Rev. Lett.}\ }\textbf {\bibinfo {volume} {100}},\ \bibinfo
  {eid} {165301} (\bibinfo {year} {2008})}\BibitemShut {NoStop}%
\bibitem [{\citenamefont {McGehee}\ \emph {et~al.}(2013)\citenamefont
  {McGehee}, \citenamefont {Kondov}, \citenamefont {Xu}, \citenamefont
  {Zirbel},\ and\ \citenamefont {DeMarco}}]{McGehee2013}%
  \BibitemOpen
  \bibfield  {author} {\bibinfo {author} {\bibfnamefont {W.~R.}\ \bibnamefont
  {McGehee}}, \bibinfo {author} {\bibfnamefont {S.~S.}\ \bibnamefont {Kondov}},
  \bibinfo {author} {\bibfnamefont {W.}~\bibnamefont {Xu}}, \bibinfo {author}
  {\bibfnamefont {J.~J.}\ \bibnamefont {Zirbel}}, \ and\ \bibinfo {author}
  {\bibfnamefont {B.}~\bibnamefont {DeMarco}},\ }\href {\doibase
  10.1103/PhysRevLett.111.145303} {\bibfield  {journal} {\bibinfo  {journal}
  {Phys. Rev. Lett.}\ }\textbf {\bibinfo {volume} {111}},\ \bibinfo {pages}
  {145303} (\bibinfo {year} {2013})}\BibitemShut {NoStop}%
\bibitem [{\citenamefont {M{\"u}ller}\ and\ \citenamefont
  {Shapiro}(2014)}]{Mueller2014}%
  \BibitemOpen
  \bibfield  {author} {\bibinfo {author} {\bibfnamefont {C.~A.}\ \bibnamefont
  {M{\"u}ller}}\ and\ \bibinfo {author} {\bibfnamefont {B.}~\bibnamefont
  {Shapiro}},\ }\href {\doibase 10.1103/PhysRevLett.113.099601} {\bibfield
  {journal} {\bibinfo  {journal} {Phys. Rev. Lett.}\ }\textbf {\bibinfo
  {volume} {113}},\ \bibinfo {pages} {099601} (\bibinfo {year}
  {2014})}\BibitemShut {NoStop}%
\bibitem [{\citenamefont {McGehee}\ \emph {et~al.}(2014)\citenamefont
  {McGehee}, \citenamefont {Kondov}, \citenamefont {Xu}, \citenamefont
  {Zirbel},\ and\ \citenamefont {DeMarco}}]{McGehee2014}%
  \BibitemOpen
  \bibfield  {author} {\bibinfo {author} {\bibfnamefont {W.~R.}\ \bibnamefont
  {McGehee}}, \bibinfo {author} {\bibfnamefont {S.~S.}\ \bibnamefont {Kondov}},
  \bibinfo {author} {\bibfnamefont {W.}~\bibnamefont {Xu}}, \bibinfo {author}
  {\bibfnamefont {J.~J.}\ \bibnamefont {Zirbel}}, \ and\ \bibinfo {author}
  {\bibfnamefont {B.}~\bibnamefont {DeMarco}},\ }\href {\doibase
  10.1103/PhysRevLett.113.099602} {\bibfield  {journal} {\bibinfo  {journal}
  {Phys. Rev. Lett.}\ }\textbf {\bibinfo {volume} {113}},\ \bibinfo {pages}
  {099602} (\bibinfo {year} {2014})}\BibitemShut {NoStop}%
\bibitem [{\citenamefont {Delande}\ and\ \citenamefont
  {Orso}(2014)}]{Delande2014}%
  \BibitemOpen
  \bibfield  {author} {\bibinfo {author} {\bibfnamefont {D.}~\bibnamefont
  {Delande}}\ and\ \bibinfo {author} {\bibfnamefont {G.}~\bibnamefont {Orso}},\
  }\href {\doibase 10.1103/PhysRevLett.113.060601} {\bibfield  {journal}
  {\bibinfo  {journal} {Phys. Rev. Lett.}\ }\textbf {\bibinfo {volume} {113}},\
  \bibinfo {pages} {060601} (\bibinfo {year} {2014})}\BibitemShut {NoStop}%
\bibitem [{\citenamefont {Basko}\ \emph {et~al.}(2006)\citenamefont {Basko},
  \citenamefont {Aleiner},\ and\ \citenamefont {Altshuler}}]{Basko2006}%
  \BibitemOpen
  \bibfield  {author} {\bibinfo {author} {\bibfnamefont {D.}~\bibnamefont
  {Basko}}, \bibinfo {author} {\bibfnamefont {I.}~\bibnamefont {Aleiner}}, \
  and\ \bibinfo {author} {\bibfnamefont {B.}~\bibnamefont {Altshuler}},\ }\href
  {\doibase 10.1016/j.aop.2005.11.014} {\bibfield  {journal} {\bibinfo
  {journal} {Ann. Phys.}\ }\textbf {\bibinfo {volume} {321}},\ \bibinfo {pages}
  {1126} (\bibinfo {year} {2006})}\BibitemShut {NoStop}%
\bibitem [{\citenamefont {Nandkishore}\ and\ \citenamefont
  {Huse}(2015)}]{Nandkishore2015}%
  \BibitemOpen
  \bibfield  {author} {\bibinfo {author} {\bibfnamefont {R.}~\bibnamefont
  {Nandkishore}}\ and\ \bibinfo {author} {\bibfnamefont {D.~A.}\ \bibnamefont
  {Huse}},\ }\href {\doibase 10.1146/annurev-conmatphys-031214-014726}
  {\bibfield  {journal} {\bibinfo  {journal} {Annu. Rev. Cond. Mat. Phys.}\
  }\textbf {\bibinfo {volume} {6}},\ \bibinfo {pages} {15} (\bibinfo {year}
  {2015})}\BibitemShut {NoStop}%
\bibitem [{\citenamefont {Altman}\ and\ \citenamefont
  {Vosk}(2015)}]{Altman2015}%
  \BibitemOpen
  \bibfield  {author} {\bibinfo {author} {\bibfnamefont {E.}~\bibnamefont
  {Altman}}\ and\ \bibinfo {author} {\bibfnamefont {R.}~\bibnamefont {Vosk}},\
  }\href {\doibase 10.1146/annurev-conmatphys-031214-014701} {\bibfield
  {journal} {\bibinfo  {journal} {Annu. Rev. Cond. Mat. Phys.}\ }\textbf
  {\bibinfo {volume} {6}},\ \bibinfo {pages} {383} (\bibinfo {year}
  {2015})}\BibitemShut {NoStop}%
\bibitem [{\citenamefont {Schreiber}\ \emph {et~al.}(2015)\citenamefont
  {Schreiber}, \citenamefont {Hodgman}, \citenamefont {Bordia}, \citenamefont
  {L{\"u}schen}, \citenamefont {Fischer}, \citenamefont {Vosk}, \citenamefont
  {Altman}, \citenamefont {Schneider},\ and\ \citenamefont
  {Bloch}}]{Schreiber2015}%
  \BibitemOpen
  \bibfield  {author} {\bibinfo {author} {\bibfnamefont {M.}~\bibnamefont
  {Schreiber}}, \bibinfo {author} {\bibfnamefont {S.~S.}\ \bibnamefont
  {Hodgman}}, \bibinfo {author} {\bibfnamefont {P.}~\bibnamefont {Bordia}},
  \bibinfo {author} {\bibfnamefont {H.~P.}\ \bibnamefont {L{\"u}schen}},
  \bibinfo {author} {\bibfnamefont {M.~H.}\ \bibnamefont {Fischer}}, \bibinfo
  {author} {\bibfnamefont {R.}~\bibnamefont {Vosk}}, \bibinfo {author}
  {\bibfnamefont {E.}~\bibnamefont {Altman}}, \bibinfo {author} {\bibfnamefont
  {U.}~\bibnamefont {Schneider}}, \ and\ \bibinfo {author} {\bibfnamefont
  {I.}~\bibnamefont {Bloch}},\ }\href {\doibase 10.1126/science.aaa7432}
  {\bibfield  {journal} {\bibinfo  {journal} {Science}\ }\textbf {\bibinfo
  {volume} {349}},\ \bibinfo {pages} {842} (\bibinfo {year}
  {2015})}\BibitemShut {NoStop}%
\bibitem [{\citenamefont {Bordia}\ \emph {et~al.}(2016)\citenamefont {Bordia},
  \citenamefont {L\"uschen}, \citenamefont {Hodgman}, \citenamefont
  {Schreiber}, \citenamefont {Bloch},\ and\ \citenamefont
  {Schneider}}]{Bordia2016}%
  \BibitemOpen
  \bibfield  {author} {\bibinfo {author} {\bibfnamefont {P.}~\bibnamefont
  {Bordia}}, \bibinfo {author} {\bibfnamefont {H.~P.}\ \bibnamefont
  {L\"uschen}}, \bibinfo {author} {\bibfnamefont {S.~S.}\ \bibnamefont
  {Hodgman}}, \bibinfo {author} {\bibfnamefont {M.}~\bibnamefont {Schreiber}},
  \bibinfo {author} {\bibfnamefont {I.}~\bibnamefont {Bloch}}, \ and\ \bibinfo
  {author} {\bibfnamefont {U.}~\bibnamefont {Schneider}},\ }\href {\doibase
  10.1103/PhysRevLett.116.140401} {\bibfield  {journal} {\bibinfo  {journal}
  {Phys. Rev. Lett.}\ }\textbf {\bibinfo {volume} {116}},\ \bibinfo {pages}
  {140401} (\bibinfo {year} {2016})}\BibitemShut {NoStop}%
\bibitem [{\citenamefont {Agarwal}\ \emph {et~al.}(2015)\citenamefont
  {Agarwal}, \citenamefont {Gopalakrishnan}, \citenamefont {Knap},
  \citenamefont {M\"uller},\ and\ \citenamefont {Demler}}]{Agarwal2015}%
  \BibitemOpen
  \bibfield  {author} {\bibinfo {author} {\bibfnamefont {K.}~\bibnamefont
  {Agarwal}}, \bibinfo {author} {\bibfnamefont {S.}~\bibnamefont
  {Gopalakrishnan}}, \bibinfo {author} {\bibfnamefont {M.}~\bibnamefont
  {Knap}}, \bibinfo {author} {\bibfnamefont {M.}~\bibnamefont {M\"uller}}, \
  and\ \bibinfo {author} {\bibfnamefont {E.}~\bibnamefont {Demler}},\ }\href
  {\doibase 10.1103/PhysRevLett.114.160401} {\bibfield  {journal} {\bibinfo
  {journal} {Phys. Rev. Lett.}\ }\textbf {\bibinfo {volume} {114}},\ \bibinfo
  {pages} {160401} (\bibinfo {year} {2015})}\BibitemShut {NoStop}%
\bibitem [{\citenamefont {Potter}\ \emph {et~al.}(2015)\citenamefont {Potter},
  \citenamefont {Vasseur},\ and\ \citenamefont {Parameswaran}}]{Potter2015}%
  \BibitemOpen
  \bibfield  {author} {\bibinfo {author} {\bibfnamefont {A.~C.}\ \bibnamefont
  {Potter}}, \bibinfo {author} {\bibfnamefont {R.}~\bibnamefont {Vasseur}}, \
  and\ \bibinfo {author} {\bibfnamefont {S.~A.}\ \bibnamefont {Parameswaran}},\
  }\href {\doibase 10.1103/PhysRevX.5.031033} {\bibfield  {journal} {\bibinfo
  {journal} {Phys. Rev. X}\ }\textbf {\bibinfo {volume} {5}},\ \bibinfo {pages}
  {031033} (\bibinfo {year} {2015})}\BibitemShut {NoStop}%
\bibitem [{\citenamefont {Vosk}\ \emph {et~al.}(2015)\citenamefont {Vosk},
  \citenamefont {Huse},\ and\ \citenamefont {Altman}}]{Vosk2015}%
  \BibitemOpen
  \bibfield  {author} {\bibinfo {author} {\bibfnamefont {R.}~\bibnamefont
  {Vosk}}, \bibinfo {author} {\bibfnamefont {D.~A.}\ \bibnamefont {Huse}}, \
  and\ \bibinfo {author} {\bibfnamefont {E.}~\bibnamefont {Altman}},\ }\href
  {\doibase 10.1103/PhysRevX.5.031032} {\bibfield  {journal} {\bibinfo
  {journal} {Phys. Rev. X}\ }\textbf {\bibinfo {volume} {5}},\ \bibinfo {pages}
  {031032} (\bibinfo {year} {2015})}\BibitemShut {NoStop}%
\bibitem [{\citenamefont {Bar~Lev}\ and\ \citenamefont
  {Reichman}(2016)}]{BarLev2016}%
  \BibitemOpen
  \bibfield  {author} {\bibinfo {author} {\bibfnamefont {Y.}~\bibnamefont
  {Bar~Lev}}\ and\ \bibinfo {author} {\bibfnamefont {D.~R.}\ \bibnamefont
  {Reichman}},\ }\href {http://stacks.iop.org/0295-5075/113/i=4/a=46001}
  {\bibfield  {journal} {\bibinfo  {journal} {Europhys. Lett.}\ }\textbf
  {\bibinfo {volume} {113}},\ \bibinfo {pages} {46001} (\bibinfo {year}
  {2016})}\BibitemShut {NoStop}%
\bibitem [{\citenamefont {{Choi}}\ \emph {et~al.}(2016)\citenamefont {{Choi}},
  \citenamefont {{Hild}}, \citenamefont {{Zeiher}}, \citenamefont
  {{Schau{\ss}}}, \citenamefont {{Rubio-Abadal}}, \citenamefont {{Yefsah}},
  \citenamefont {{Khemani}}, \citenamefont {{Huse}}, \citenamefont {{Bloch}},\
  and\ \citenamefont {{Gross}}}]{Choi2016}%
  \BibitemOpen
  \bibfield  {author} {\bibinfo {author} {\bibfnamefont {J.-y.}\ \bibnamefont
  {{Choi}}}, \bibinfo {author} {\bibfnamefont {S.}~\bibnamefont {{Hild}}},
  \bibinfo {author} {\bibfnamefont {J.}~\bibnamefont {{Zeiher}}}, \bibinfo
  {author} {\bibfnamefont {P.}~\bibnamefont {{Schau{\ss}}}}, \bibinfo {author}
  {\bibfnamefont {A.}~\bibnamefont {{Rubio-Abadal}}}, \bibinfo {author}
  {\bibfnamefont {T.}~\bibnamefont {{Yefsah}}}, \bibinfo {author}
  {\bibfnamefont {V.}~\bibnamefont {{Khemani}}}, \bibinfo {author}
  {\bibfnamefont {D.~A.}\ \bibnamefont {{Huse}}}, \bibinfo {author}
  {\bibfnamefont {I.}~\bibnamefont {{Bloch}}}, \ and\ \bibinfo {author}
  {\bibfnamefont {C.}~\bibnamefont {{Gross}}},\ }\href@noop {} {\bibfield
  {journal} {\bibinfo  {journal} {ArXiv e-prints}\ } (\bibinfo {year}
  {2016})},\ \Eprint {http://arxiv.org/abs/1604.04178} {arXiv:1604.04178}
  \BibitemShut {NoStop}%
\bibitem [{\citenamefont {Luitz}\ \emph {et~al.}(2016)\citenamefont {Luitz},
  \citenamefont {Laflorencie},\ and\ \citenamefont {Alet}}]{Luitz2016}%
  \BibitemOpen
  \bibfield  {author} {\bibinfo {author} {\bibfnamefont {D.~J.}\ \bibnamefont
  {Luitz}}, \bibinfo {author} {\bibfnamefont {N.}~\bibnamefont {Laflorencie}},
  \ and\ \bibinfo {author} {\bibfnamefont {F.}~\bibnamefont {Alet}},\ }\href
  {\doibase 10.1103/PhysRevB.93.060201} {\bibfield  {journal} {\bibinfo
  {journal} {Phys. Rev. B}\ }\textbf {\bibinfo {volume} {93}},\ \bibinfo
  {pages} {060201} (\bibinfo {year} {2016})}\BibitemShut {NoStop}%
\bibitem [{\citenamefont {Piraud}\ \emph {et~al.}(2012)\citenamefont {Piraud},
  \citenamefont {Pezz{\'e}},\ and\ \citenamefont
  {Sanchez-Palencia}}]{Piraud2012}%
  \BibitemOpen
  \bibfield  {author} {\bibinfo {author} {\bibfnamefont {M.}~\bibnamefont
  {Piraud}}, \bibinfo {author} {\bibfnamefont {L.}~\bibnamefont {Pezz{\'e}}}, \
  and\ \bibinfo {author} {\bibfnamefont {L.}~\bibnamefont {Sanchez-Palencia}},\
  }\href {http://stacks.iop.org/0295-5075/99/i=5/a=50003} {\bibfield  {journal}
  {\bibinfo  {journal} {Europhys. Lett.}\ }\textbf {\bibinfo {volume} {99}},\
  \bibinfo {pages} {50003} (\bibinfo {year} {2012})}\BibitemShut {NoStop}%
\bibitem [{\citenamefont {Piraud}\ \emph {et~al.}(2013)\citenamefont {Piraud},
  \citenamefont {Pezz{\'e}},\ and\ \citenamefont
  {Sanchez-Palencia}}]{Piraud2013}%
  \BibitemOpen
  \bibfield  {author} {\bibinfo {author} {\bibfnamefont {M.}~\bibnamefont
  {Piraud}}, \bibinfo {author} {\bibfnamefont {L.}~\bibnamefont {Pezz{\'e}}}, \
  and\ \bibinfo {author} {\bibfnamefont {L.}~\bibnamefont {Sanchez-Palencia}},\
  }\href {http://stacks.iop.org/1367-2630/15/i=7/a=075007} {\bibfield
  {journal} {\bibinfo  {journal} {New J. Phys.}\ }\textbf {\bibinfo {volume}
  {15}},\ \bibinfo {pages} {075007} (\bibinfo {year} {2013})}\BibitemShut
  {NoStop}%
\bibitem [{\citenamefont {Piraud}\ \emph {et~al.}(2014)\citenamefont {Piraud},
  \citenamefont {Sanchez-Palencia},\ and\ \citenamefont {van
  Tiggelen}}]{Piraud2014}%
  \BibitemOpen
  \bibfield  {author} {\bibinfo {author} {\bibfnamefont {M.}~\bibnamefont
  {Piraud}}, \bibinfo {author} {\bibfnamefont {L.}~\bibnamefont
  {Sanchez-Palencia}}, \ and\ \bibinfo {author} {\bibfnamefont
  {B.}~\bibnamefont {van Tiggelen}},\ }\href {\doibase
  10.1103/PhysRevA.90.063639} {\bibfield  {journal} {\bibinfo  {journal} {Phys.
  Rev. A}\ }\textbf {\bibinfo {volume} {90}},\ \bibinfo {pages} {063639}
  (\bibinfo {year} {2014})}\BibitemShut {NoStop}%
\bibitem [{Note1()}]{Note1}%
  \BibitemOpen
  \bibinfo {note} {Strictly speaking, 3D speckle potentials created by a
  superposition of monochromatic beams from ideal rectangular or circular
  apertures, for which a white-noise limit does not exist, are not generic in
  this sense \cite {Kuhn2007,Shapiro2012}. This caveat only affects the
  estimate \protect \textup {\hbox {\mathsurround \z@ \protect \normalfont
  (\ignorespaces \ref {width}\unskip \@@italiccorr )}} for the critical scale
  $W$, see \cite {Note3}, but does not invalidate our general
  analysis.}\BibitemShut {Stop}%
\bibitem [{\citenamefont {Yedjour}\ and\ \citenamefont
  {Tiggelen}(2010)}]{Yedjour2010}%
  \BibitemOpen
  \bibfield  {author} {\bibinfo {author} {\bibfnamefont {A.}~\bibnamefont
  {Yedjour}}\ and\ \bibinfo {author} {\bibfnamefont {B.~A.}\ \bibnamefont
  {Tiggelen}},\ }\href {\doibase 10.1140/epjd/e2010-00141-5} {\bibfield
  {journal} {\bibinfo  {journal} {Eur. Phys. J. D}\ }\textbf {\bibinfo {volume}
  {59}},\ \bibinfo {pages} {249} (\bibinfo {year} {2010})}\BibitemShut
  {NoStop}%
\bibitem [{Note2()}]{Note2}%
  \BibitemOpen
  \bibinfo {note} {A simple on-shell approximation for the spectral function
  puts $E_\protect \text {c}$ above sea-level \cite {Kuhn2007}, but this value
  is `red-shifted' below sea-level by the real part of the single-particle
  self-energy \cite {Piraud2013}.}\BibitemShut {Stop}%
\bibitem [{Note3()}]{Note3}%
  \BibitemOpen
  \bibinfo {note} {The estimate \protect \textup {\hbox {\mathsurround \z@
  \protect \normalfont (\ignorespaces \ref {width}\unskip \@@italiccorr )}}
  holds for generic disorder. In non-generic 3D speckle with an infrared
  divergence of the 3D pair correlator $\protect \mathaccentV
  {tilde}07EC(\protect \mathbf {k})$ \cite {Kuhn2007}, one rather has an
  energy-dependent mean free path $l_E\sim k_E\zeta ^2/\eta ^2$, such that
  $W\sim \eta ^2 E_\zeta = \eta V_0$, even larger than \protect \textup {\hbox
  {\mathsurround \z@ \protect \normalfont (\ignorespaces \ref {width}\unskip
  \@@italiccorr )}}.}\BibitemShut {Stop}%
\bibitem [{\citenamefont {Trappe}\ \emph {et~al.}(2015)\citenamefont {Trappe},
  \citenamefont {Delande},\ and\ \citenamefont {M\"{u}ller}}]{Trappe2015}%
  \BibitemOpen
  \bibfield  {author} {\bibinfo {author} {\bibfnamefont {M.~I.}\ \bibnamefont
  {Trappe}}, \bibinfo {author} {\bibfnamefont {D.}~\bibnamefont {Delande}}, \
  and\ \bibinfo {author} {\bibfnamefont {C.~A.}\ \bibnamefont {M\"{u}ller}},\
  }\href {http://stacks.iop.org/1751-8121/48/i=24/a=245102} {\bibfield
  {journal} {\bibinfo  {journal} {J. Phys. A: Math. Theo.}\ }\textbf {\bibinfo
  {volume} {48}},\ \bibinfo {pages} {245102} (\bibinfo {year}
  {2015})}\BibitemShut {NoStop}%
\bibitem [{\citenamefont {Lifshitz}(1964)}]{Lifshitz1964}%
  \BibitemOpen
  \bibfield  {author} {\bibinfo {author} {\bibfnamefont {I.~M.}\ \bibnamefont
  {Lifshitz}},\ }\href {\doibase 10.1080/00018736400101061} {\bibfield
  {journal} {\bibinfo  {journal} {Adv. Phys.}\ }\textbf {\bibinfo {volume}
  {13}},\ \bibinfo {pages} {483} (\bibinfo {year} {1964})}\BibitemShut
  {NoStop}%
\bibitem [{\citenamefont {Lifshitz}\ \emph {et~al.}(1988)\citenamefont
  {Lifshitz}, \citenamefont {Gredeskul},\ and\ \citenamefont
  {Pastur}}]{Lifshitz1988}%
  \BibitemOpen
  \bibfield  {author} {\bibinfo {author} {\bibfnamefont {I.~M.}\ \bibnamefont
  {Lifshitz}}, \bibinfo {author} {\bibfnamefont {S.~A.}\ \bibnamefont
  {Gredeskul}}, \ and\ \bibinfo {author} {\bibfnamefont {L.~A.}\ \bibnamefont
  {Pastur}},\ }\href@noop {} {\emph {\bibinfo {title} {Introduction to the
  Theory of Disordered Systems}}}\ (\bibinfo  {publisher} {Wiley, New York},\
  \bibinfo {year} {1988})\BibitemShut {NoStop}%
\bibitem [{\citenamefont {Pasek}\ \emph {et~al.}(2015)\citenamefont {Pasek},
  \citenamefont {Zhao}, \citenamefont {Delande},\ and\ \citenamefont
  {Orso}}]{Pasek2015}%
  \BibitemOpen
  \bibfield  {author} {\bibinfo {author} {\bibfnamefont {M.}~\bibnamefont
  {Pasek}}, \bibinfo {author} {\bibfnamefont {Z.}~\bibnamefont {Zhao}},
  \bibinfo {author} {\bibfnamefont {D.}~\bibnamefont {Delande}}, \ and\
  \bibinfo {author} {\bibfnamefont {G.}~\bibnamefont {Orso}},\ }\href {\doibase
  10.1103/PhysRevA.92.053618} {\bibfield  {journal} {\bibinfo  {journal} {Phys.
  Rev. A}\ }\textbf {\bibinfo {volume} {92}},\ \bibinfo {pages} {053618}
  (\bibinfo {year} {2015})}\BibitemShut {NoStop}%
\bibitem [{\citenamefont {Slevin}\ and\ \citenamefont
  {Ohtsuki}(1999)}]{Slevin1999}%
  \BibitemOpen
  \bibfield  {author} {\bibinfo {author} {\bibfnamefont {K.}~\bibnamefont
  {Slevin}}\ and\ \bibinfo {author} {\bibfnamefont {T.}~\bibnamefont
  {Ohtsuki}},\ }\href {\doibase 10.1103/PhysRevLett.82.382} {\bibfield
  {journal} {\bibinfo  {journal} {Phys. Rev. Lett.}\ }\textbf {\bibinfo
  {volume} {82}},\ \bibinfo {pages} {382} (\bibinfo {year} {1999})}\BibitemShut
  {NoStop}%
\bibitem [{\citenamefont {Rodriguez}\ \emph {et~al.}(2010)\citenamefont
  {Rodriguez}, \citenamefont {Vasquez}, \citenamefont {Slevin},\ and\
  \citenamefont {R\"omer}}]{Rodriguez2010}%
  \BibitemOpen
  \bibfield  {author} {\bibinfo {author} {\bibfnamefont {A.}~\bibnamefont
  {Rodriguez}}, \bibinfo {author} {\bibfnamefont {L.~J.}\ \bibnamefont
  {Vasquez}}, \bibinfo {author} {\bibfnamefont {K.}~\bibnamefont {Slevin}}, \
  and\ \bibinfo {author} {\bibfnamefont {R.~A.}\ \bibnamefont {R\"omer}},\
  }\href {\doibase 10.1103/PhysRevLett.105.046403} {\bibfield  {journal}
  {\bibinfo  {journal} {Phys. Rev. Lett.}\ }\textbf {\bibinfo {volume} {105}},\
  \bibinfo {pages} {046403} (\bibinfo {year} {2010})}\BibitemShut {NoStop}%
\bibitem [{\citenamefont {Lopez}\ \emph {et~al.}(2012)\citenamefont {Lopez},
  \citenamefont {Cl\'ement}, \citenamefont {Szriftgiser}, \citenamefont
  {Garreau},\ and\ \citenamefont {Delande}}]{Lopez2012}%
  \BibitemOpen
  \bibfield  {author} {\bibinfo {author} {\bibfnamefont {M.}~\bibnamefont
  {Lopez}}, \bibinfo {author} {\bibfnamefont {J.-F.}\ \bibnamefont
  {Cl\'ement}}, \bibinfo {author} {\bibfnamefont {P.}~\bibnamefont
  {Szriftgiser}}, \bibinfo {author} {\bibfnamefont {J.~C.}\ \bibnamefont
  {Garreau}}, \ and\ \bibinfo {author} {\bibfnamefont {D.}~\bibnamefont
  {Delande}},\ }\href {\doibase 10.1103/PhysRevLett.108.095701} {\bibfield
  {journal} {\bibinfo  {journal} {Phys. Rev. Lett.}\ }\textbf {\bibinfo
  {volume} {108}},\ \bibinfo {pages} {095701} (\bibinfo {year}
  {2012})}\BibitemShut {NoStop}%
\bibitem [{\citenamefont {Slevin}\ and\ \citenamefont
  {Ohtsuki}(2014)}]{Slevin2014}%
  \BibitemOpen
  \bibfield  {author} {\bibinfo {author} {\bibfnamefont {K.}~\bibnamefont
  {Slevin}}\ and\ \bibinfo {author} {\bibfnamefont {T.}~\bibnamefont
  {Ohtsuki}},\ }\href {http://stacks.iop.org/1367-2630/16/i=1/a=015012}
  {\bibfield  {journal} {\bibinfo  {journal} {New J. Phys.}\ }\textbf {\bibinfo
  {volume} {16}},\ \bibinfo {pages} {015012} (\bibinfo {year}
  {2014})}\BibitemShut {NoStop}%
\bibitem [{\citenamefont {Wegner}(1976)}]{Wegner1976}%
  \BibitemOpen
  \bibfield  {author} {\bibinfo {author} {\bibfnamefont {F.~J.}\ \bibnamefont
  {Wegner}},\ }\href {\doibase 10.1007/BF01315248} {\bibfield  {journal}
  {\bibinfo  {journal} {Z. Phys. B}\ }\textbf {\bibinfo {volume} {25}},\
  \bibinfo {pages} {327} (\bibinfo {year} {1976})}\BibitemShut {NoStop}%
\bibitem [{\citenamefont {Kramer}\ and\ \citenamefont
  {MacKinnon}(1993)}]{Kramer1993}%
  \BibitemOpen
  \bibfield  {author} {\bibinfo {author} {\bibfnamefont {B.}~\bibnamefont
  {Kramer}}\ and\ \bibinfo {author} {\bibfnamefont {A.}~\bibnamefont
  {MacKinnon}},\ }\href@noop {} {\bibfield  {journal} {\bibinfo  {journal}
  {Rep. Progr. Phys.}\ }\textbf {\bibinfo {volume} {56}},\ \bibinfo {pages}
  {1469} (\bibinfo {year} {1993})}\BibitemShut {NoStop}%
\bibitem [{\citenamefont {Soukoulis}\ and\ \citenamefont
  {Economou}(1999)}]{Soukoulis1999}%
  \BibitemOpen
  \bibfield  {author} {\bibinfo {author} {\bibfnamefont {C.}~\bibnamefont
  {Soukoulis}}\ and\ \bibinfo {author} {\bibfnamefont {E.}~\bibnamefont
  {Economou}},\ }\href@noop {} {\bibfield  {journal} {\bibinfo  {journal}
  {Waves Rand. Media}\ }\textbf {\bibinfo {volume} {9}},\ \bibinfo {pages}
  {255} (\bibinfo {year} {1999})}\BibitemShut {NoStop}%
\bibitem [{\citenamefont {Efetov}(1997)}]{Efetov1997}%
  \BibitemOpen
  \bibfield  {author} {\bibinfo {author} {\bibfnamefont {K.}~\bibnamefont
  {Efetov}},\ }\href@noop {} {\emph {\bibinfo {title} {{Supersymmetry in
  disorder and chaos}}}}\ (\bibinfo  {publisher} {Cambridge Univ. Press},\
  \bibinfo {year} {1997})\BibitemShut {NoStop}%
\bibitem [{\citenamefont {Ohtsuki}\ and\ \citenamefont
  {Kawarabayashi}(1997)}]{Ohtsuki1997}%
  \BibitemOpen
  \bibfield  {author} {\bibinfo {author} {\bibfnamefont {T.}~\bibnamefont
  {Ohtsuki}}\ and\ \bibinfo {author} {\bibfnamefont {T.}~\bibnamefont
  {Kawarabayashi}},\ }\href {\doibase 10.1143/JPSJ.66.314} {\bibfield
  {journal} {\bibinfo  {journal} {J. Phys. Soc. Jpn.}\ }\textbf {\bibinfo
  {volume} {66}},\ \bibinfo {pages} {314} (\bibinfo {year} {1997})}\BibitemShut
  {NoStop}%
\bibitem [{\citenamefont {Lemari{\'e}}\ \emph {et~al.}(2010)\citenamefont
  {Lemari{\'e}}, \citenamefont {Lignier}, \citenamefont {Delande},
  \citenamefont {Szriftgiser},\ and\ \citenamefont {Garreau}}]{Lemarie2010}%
  \BibitemOpen
  \bibfield  {author} {\bibinfo {author} {\bibfnamefont {G.}~\bibnamefont
  {Lemari{\'e}}}, \bibinfo {author} {\bibfnamefont {H.}~\bibnamefont
  {Lignier}}, \bibinfo {author} {\bibfnamefont {D.}~\bibnamefont {Delande}},
  \bibinfo {author} {\bibfnamefont {P.}~\bibnamefont {Szriftgiser}}, \ and\
  \bibinfo {author} {\bibfnamefont {J.~C.}\ \bibnamefont {Garreau}},\ }\href
  {\doibase 10.1103/PhysRevLett.105.090601} {\bibfield  {journal} {\bibinfo
  {journal} {Phys. Rev. Lett.}\ }\textbf {\bibinfo {volume} {105}},\ \bibinfo
  {pages} {090601} (\bibinfo {year} {2010})}\BibitemShut {NoStop}%
\bibitem [{\citenamefont {Shapiro}(1982)}]{Shapiro1982}%
  \BibitemOpen
  \bibfield  {author} {\bibinfo {author} {\bibfnamefont {B.}~\bibnamefont
  {Shapiro}},\ }\href {\doibase 10.1103/PhysRevB.25.4266} {\bibfield  {journal}
  {\bibinfo  {journal} {Phys. Rev. B}\ }\textbf {\bibinfo {volume} {25}},\
  \bibinfo {pages} {4266} (\bibinfo {year} {1982})}\BibitemShut {NoStop}%
\bibitem [{\citenamefont {Lemari\'e}\ \emph {et~al.}(2009)\citenamefont
  {Lemari\'e}, \citenamefont {Chab\'e}, \citenamefont {Szriftgiser},
  \citenamefont {Garreau}, \citenamefont {Gr\'emaud},\ and\ \citenamefont
  {Delande}}]{Lemarie2009}%
  \BibitemOpen
  \bibfield  {author} {\bibinfo {author} {\bibfnamefont {G.}~\bibnamefont
  {Lemari\'e}}, \bibinfo {author} {\bibfnamefont {J.}~\bibnamefont {Chab\'e}},
  \bibinfo {author} {\bibfnamefont {P.}~\bibnamefont {Szriftgiser}}, \bibinfo
  {author} {\bibfnamefont {J.~C.}\ \bibnamefont {Garreau}}, \bibinfo {author}
  {\bibfnamefont {B.}~\bibnamefont {Gr\'emaud}}, \ and\ \bibinfo {author}
  {\bibfnamefont {D.}~\bibnamefont {Delande}},\ }\href {\doibase
  10.1103/PhysRevA.80.043626} {\bibfield  {journal} {\bibinfo  {journal} {Phys.
  Rev. A}\ }\textbf {\bibinfo {volume} {80}},\ \bibinfo {pages} {043626}
  (\bibinfo {year} {2009})}\BibitemShut {NoStop}%
\bibitem [{\citenamefont {Karpiuk}\ \emph {et~al.}(2012)\citenamefont
  {Karpiuk}, \citenamefont {Cherroret}, \citenamefont {Lee}, \citenamefont
  {Gr{\'e}maud}, \citenamefont {M{\"u}ller},\ and\ \citenamefont
  {Miniatura}}]{Karpiuk2012}%
  \BibitemOpen
  \bibfield  {author} {\bibinfo {author} {\bibfnamefont {T.}~\bibnamefont
  {Karpiuk}}, \bibinfo {author} {\bibfnamefont {N.}~\bibnamefont {Cherroret}},
  \bibinfo {author} {\bibfnamefont {K.~L.}\ \bibnamefont {Lee}}, \bibinfo
  {author} {\bibfnamefont {B.}~\bibnamefont {Gr{\'e}maud}}, \bibinfo {author}
  {\bibfnamefont {C.~A.}\ \bibnamefont {M{\"u}ller}}, \ and\ \bibinfo {author}
  {\bibfnamefont {C.}~\bibnamefont {Miniatura}},\ }\href {\doibase
  10.1103/PhysRevLett.109.190601} {\bibfield  {journal} {\bibinfo  {journal}
  {Phys. Rev. Lett.}\ }\textbf {\bibinfo {volume} {109}},\ \bibinfo {pages}
  {190601} (\bibinfo {year} {2012})}\BibitemShut {NoStop}%
\bibitem [{\citenamefont {Lee}\ \emph {et~al.}(2014)\citenamefont {Lee},
  \citenamefont {Gr{\'e}maud},\ and\ \citenamefont {Miniatura}}]{Lee2014}%
  \BibitemOpen
  \bibfield  {author} {\bibinfo {author} {\bibfnamefont {K.~L.}\ \bibnamefont
  {Lee}}, \bibinfo {author} {\bibfnamefont {B.}~\bibnamefont {Gr{\'e}maud}}, \
  and\ \bibinfo {author} {\bibfnamefont {C.}~\bibnamefont {Miniatura}},\ }\href
  {\doibase 10.1103/PhysRevA.90.043605} {\bibfield  {journal} {\bibinfo
  {journal} {Phys. Rev. A}\ }\textbf {\bibinfo {volume} {90}},\ \bibinfo
  {pages} {043605} (\bibinfo {year} {2014})}\BibitemShut {NoStop}%
\bibitem [{\citenamefont {Ghosh}\ \emph {et~al.}(2014)\citenamefont {Ghosh},
  \citenamefont {Cherroret}, \citenamefont {Gr{\'e}maud}, \citenamefont
  {Miniatura},\ and\ \citenamefont {Delande}}]{Ghosh2014}%
  \BibitemOpen
  \bibfield  {author} {\bibinfo {author} {\bibfnamefont {S.}~\bibnamefont
  {Ghosh}}, \bibinfo {author} {\bibfnamefont {N.}~\bibnamefont {Cherroret}},
  \bibinfo {author} {\bibfnamefont {B.}~\bibnamefont {Gr{\'e}maud}}, \bibinfo
  {author} {\bibfnamefont {C.}~\bibnamefont {Miniatura}}, \ and\ \bibinfo
  {author} {\bibfnamefont {D.}~\bibnamefont {Delande}},\ }\href {\doibase
  10.1103/PhysRevA.90.063602} {\bibfield  {journal} {\bibinfo  {journal} {Phys.
  Rev. A}\ }\textbf {\bibinfo {volume} {90}},\ \bibinfo {pages} {063602}
  (\bibinfo {year} {2014})}\BibitemShut {NoStop}%
\bibitem [{\citenamefont {Micklitz}\ \emph {et~al.}(2014)\citenamefont
  {Micklitz}, \citenamefont {M{\"u}ller},\ and\ \citenamefont
  {Altland}}]{Micklitz2014}%
  \BibitemOpen
  \bibfield  {author} {\bibinfo {author} {\bibfnamefont {T.}~\bibnamefont
  {Micklitz}}, \bibinfo {author} {\bibfnamefont {C.~A.}\ \bibnamefont
  {M{\"u}ller}}, \ and\ \bibinfo {author} {\bibfnamefont {A.}~\bibnamefont
  {Altland}},\ }\href {\doibase 10.1103/PhysRevLett.112.110602} {\bibfield
  {journal} {\bibinfo  {journal} {Phys. Rev. Lett.}\ }\textbf {\bibinfo
  {volume} {112}},\ \bibinfo {pages} {110602} (\bibinfo {year}
  {2014})}\BibitemShut {NoStop}%
\bibitem [{\citenamefont {Micklitz}\ \emph {et~al.}(2015)\citenamefont
  {Micklitz}, \citenamefont {M{\"u}ller},\ and\ \citenamefont
  {Altland}}]{Micklitz2015}%
  \BibitemOpen
  \bibfield  {author} {\bibinfo {author} {\bibfnamefont {T.}~\bibnamefont
  {Micklitz}}, \bibinfo {author} {\bibfnamefont {C.~A.}\ \bibnamefont
  {M{\"u}ller}}, \ and\ \bibinfo {author} {\bibfnamefont {A.}~\bibnamefont
  {Altland}},\ }\href {\doibase 10.1103/PhysRevB.91.064203} {\bibfield
  {journal} {\bibinfo  {journal} {Phys. Rev. B}\ }\textbf {\bibinfo {volume}
  {91}},\ \bibinfo {pages} {064203} (\bibinfo {year} {2015})}\BibitemShut
  {NoStop}%
\bibitem [{\citenamefont {Cherroret}\ \emph {et~al.}(2014)\citenamefont
  {Cherroret}, \citenamefont {Vermersch}, \citenamefont {Garreau},\ and\
  \citenamefont {Delande}}]{Cherroret2014}%
  \BibitemOpen
  \bibfield  {author} {\bibinfo {author} {\bibfnamefont {N.}~\bibnamefont
  {Cherroret}}, \bibinfo {author} {\bibfnamefont {B.}~\bibnamefont
  {Vermersch}}, \bibinfo {author} {\bibfnamefont {J.~C.}\ \bibnamefont
  {Garreau}}, \ and\ \bibinfo {author} {\bibfnamefont {D.}~\bibnamefont
  {Delande}},\ }\href {\doibase 10.1103/PhysRevLett.112.170603} {\bibfield
  {journal} {\bibinfo  {journal} {Phys. Rev. Lett.}\ }\textbf {\bibinfo
  {volume} {112}},\ \bibinfo {pages} {170603} (\bibinfo {year}
  {2014})}\BibitemShut {NoStop}%
\end{thebibliography}%

\end{document}